\documentclass[twocolumn]{aastex631}

\usepackage{placeins}
\usepackage{savesym}
\savesymbol{tablenum}
\usepackage[sicmds,freestanding]{hepunits}
\restoresymbol{SIX}{tablenum}
\usepackage{amssymb,mathtools,bm}
\usepackage{graphicx, color}
\usepackage[dvipsnames]{xcolor}
\usepackage{float}
\usepackage{needspace}
\usepackage{xspace}
\usepackage{filecontents}
\usepackage{multirow}
 \usepackage{hyperref}
\hypersetup{
    colorlinks=true,       
    linkcolor=blue,        
    citecolor=blue,        
    filecolor=magenta,     
    urlcolor=blue          
}

\usepackage{slashed}

\usepackage{url}
\usepackage[utf8]{inputenc}
\usepackage[english]{babel}
\usepackage{mathrsfs}
\usepackage{rotating}
\usepackage{longtable}

\newlength{\dhatheight}

\newcommand{\osim}{\ensuremath{\mathord{\sim}}}
\newcommand{\Msun}{\ensuremath{\mathrm{M}_{\odot}}}
\newcommand{\Lsun}{\ensuremath{\mathrm{L}_{\odot}}}
\newcommand{\sigmaLOS}{\ensuremath{\sigma_{\mathrm{\scriptscriptstyle LOS}}}}
\newcommand{\Mstar}{\ensuremath{M_{\star}}}
\newcommand{\Mpeak}{\ensuremath{M_\mathrm{peak}}}
\newcommand{\Mvir}{\ensuremath{M_\mathrm{vir}}}
\newcommand{\Mhalf}{\ensuremath{M_{1/2}}}
\newcommand{\rhalf}{\ensuremath{r_{1/2}}}
\newcommand{\vmax}{\ensuremath{v_\mathrm{max}}}
\newcommand{\rmax}{\ensuremath{r_\mathrm{max}}}

\newcommand{\kms}{\ensuremath{\mathrm{km}~\mathrm{s}^{-1}}}
\newcommand{\Junits}{\ensuremath{\GeV{}^2~\cm^{-5}~\steradian{}}}
\newcommand{\SatGen}{\texttt{SatGen}}

\defcitealias{Part_I}{Paper~I}
\newcommand{\paperone}{\citetalias{Part_I}}
\defcitealias{Fattahi:2018ioj}{Fattahi+18}
\newcommand{\Fattahi}{\citetalias{Fattahi:2018ioj}}
\defcitealias{2024arXiv240815214K}{Kim+24}
\newcommand{\Kim}{\citetalias{2024arXiv240815214K}}
\defcitealias{2023ApJ...956....6D}{Danieli+23}
\newcommand{\Danieli}{\citetalias{2023ApJ...956....6D}}
\defcitealias{2018MNRAS.477.1822M}{Moster+18}
\newcommand{\Moster}{\citetalias{2018MNRAS.477.1822M}}
\defcitealias{Wolf:2009tu}{Wolf+10}
\newcommand{\Wolf}{\citetalias{Wolf:2009tu}}
\defcitealias{Errani_2018}{Errani+18}

\defcitealias{Diemer:2018vmz}{Diemer+19}

\defcitealias{Zhao:2008wd}{Zhao+09}

\newcommand{\es}[2] {\begin{equation} \label{#1} \begin{split} #2 \end{split} \end{equation}}

\makeatletter
\let\frontmatter@title@above=\relax
\makeatother

\begin{document}

\input{aastex-deluxetable-spacing-fix}

\title{
Semi-analytic Inference of Satellite Densities in the Cold Dark Matter Model\\ \vspace{0.05in}
{\small Part II. Implications for Dark Matter Indirect Detection Constraints}
}

\author{Kailash Raman}
\affiliation{Leinweber Institute for Theoretical Physics, University of California, Berkeley, CA 94720, U.S.A.}
\affiliation{Theoretical Physics Group, Lawrence Berkeley National Laboratory, Berkeley, CA 94720, U.S.A.}

\author{Dylan Folsom}
\affiliation{Department of Physics, Princeton University, Princeton, NJ 08544, U.S.A.}

\author{Manoj Kaplinghat}
\affiliation{Department of Physics and Astronomy, University of California - Irvine, Irvine, CA 92697, U.S.A.}

\author{Mariangela Lisanti}
\affiliation{Department of Physics, Princeton University, Princeton, NJ 08544, U.S.A.}
\affiliation{Center for Computational Astrophysics, Flatiron Institute, New York, NY 10010, U.S.A.}

\author{Benjamin R. Safdi}
\affiliation{Leinweber Institute for Theoretical Physics, University of California, Berkeley, CA 94720, U.S.A.}
\affiliation{Theoretical Physics Group, Lawrence Berkeley National Laboratory, Berkeley, CA 94720, U.S.A.}

\date{\today}

\begin{abstract}
Dwarf galaxies provide excellent targets to search for signals of dark matter annihilation or decay. 
Using a calibrated semi-analytic model and the latest stellar kinematic data (presented in Part I), this paper updates the astrophysical $J$-factors for the Milky Way's dwarf spheroidal galaxies. We infer the probability distributions for the $J$-factors of 39 dwarfs by conditioning a population of subhalos, generated with the \SatGen{} semi-analytic satellite model, on either a dwarf's kinematically determined dynamical mass or its stellar mass. 
We also compute the $J$-factors using the Jeans equation with updated stellar kinematics and priors that incorporate varying degrees of \SatGen{} information. We use the computed $J$-factors to recast existing limits from Fermi-LAT data on the annihilation cross section. Our main result is that variations in the $J$-factors computed using the Jeans analysis introduce a factor of 2--4 uncertainty into the inferred limits on the cross section.
We argue that a cosmologically informed prior is a motivated choice that excludes thermal relic annihilation cross sections to $b\bar{b}$ below about 70~\GeV{}. For comparison, the more commonly used priors, which can lead to unphysical halo parameters, exclude masses below 130~\GeV{} at 95\% confidence level. 
We also show that the highest--$J$-factor dwarfs are spatially extended, approximately one degree on the sky, which challenges the validity of the point-source approximation adopted in many analyses of Fermi data. Finally, based on the semi-analytic model and the kinematic data, the highest $J$-factor halos have already been discovered, suggesting future ultra-faint discoveries are unlikely to substantially strengthen limits on annihilating dark matter.
\end{abstract} 

\Needspace{4\baselineskip}
\section{Introduction}

Ultra-faint dwarf galaxies~(UFDs) are prime targets for dark matter~(DM) annihilation and decay searches across the electromagnetic spectrum. This is particularly true in the gamma-ray band, where instruments such as the Fermi Large Area Telescope~(LAT;~\citealt{Fermi-LAT:2010cni,Fermi-LAT:2011vow,Fermi-LAT:2013sme,Fermi-LAT:2015att,Fermi-LAT:2015ycq,Geringer-Sameth:2014qqa,Fermi-LAT:2016uux,Calore:2018sdx,Hoof:2018hyn,McDaniel:2023bju}), the High Energy Stereoscopic System~\citep{Rinchiuso:2019etv,HESS:2020zwn}, and the upcoming Cherenkov Telescope Array Observatory~\citep{Lefranc:2016dgx}, among others~\citep{VERITAS:2017tif,HAWC:2017mfa,HAWC:2019jvm}, provide some of the most stringent constraints on the thermal relic DM annihilation cross section. However, the interpretation of these astrophysical signals depends sensitively on the modeling of the UFD DM density profiles. This work explores the primary systematic uncertainties inherent in these profile determinations and evaluates their implications for current bounds.

The DM annihilation signal is proportional to the line-of-sight (LOS) integral of the DM density squared, which is called the astrophysical $J$-factor~\citep{Bertone:2005xv,Lisanti:2016jxe,Slatyer:2017sev,Hooper:2018kfv,Safdi:2022xkm}.\footnote{For DM decays, the signal is proportional to the LOS integral of the DM density profile, which is called the $D$-factor.} The $J$-factor of a dwarf galaxy is conventionally inferred from
the kinematics of its member stars. The LOS velocity dispersion is
related through the spherical Jeans equation to the parameters of an assumed DM
density profile, typically a (generalized) Navarro--Frenk--White profile~(NFW; \citealt{Navarro:1995iw,Navarro:1996gj}), which are then constrained
in a Bayesian analysis~\citep{Strigari:2007at, Geringer-Sameth:2014yza,
Bonnivard:2015xpq, Pace:2018tin, McDaniel:2023bju}. For systems without a
dedicated kinematic analysis, the $J$-factor is sometimes estimated from
empirical scaling relations that predict it from global dwarf properties such as
distance, half-light radius, and velocity dispersion~\citep{Pace:2018tin}. Either
approach becomes increasingly uncertain in the ultra-faint regime, where the small
number of spectroscopically confirmed members leaves the inferred profile---and
thus the $J$-factor---sensitive to the adopted halo-parameter priors~\citep[though see][]{Ando:2020yyk, Horigome:2022gge}, the assumed
velocity anisotropy, and the treatment of stellar membership and unresolved
binaries~\citep{Bonnivard:2015vua}. 

In this work, we employ several principled approaches to inferring dwarf $J$-factors and use the variance between these constructions as an estimate of the current systematic uncertainty on the annihilation limits.  This paper is the second in a series; our methods follow from~\citet[][hereafter \paperone]{Part_I}, which present a semi-analytic formalism for inferring the halo properties of the Milky Way's~(MW's) dwarf galaxies.  In particular, that work utilized  the \SatGen{} satellite generator~\citep{Jiang:2020rdj,Green:2021vkf} to predict the unobserved halo properties of the MW dwarf galaxies using two different methods: one based on dwarf stellar kinematics and the other based on the stellar-to-halo mass~(SHMR) relation.  In this work, we compute the $J$-factor for nearby UFDs using \SatGen{} with both the kinematic and SHMR approaches, including updated observational data, and present a variant of the traditional Jeans analysis framework with priors conditioned on the \SatGen{} expectation.    

To highlight the impact of the $J$-factor results, we illustrate how the different inference methods affect the DM annihilation limit with Fermi gamma-ray data towards nearby UFDs.  To this end, we recast the velocity-averaged annihilation cross-section limits of
\citet{Circiello:2026inp}, taking advantage of the
publicly available likelihood profiles from their 16-year Fermi-LAT
analysis, which can be reweighted to an arbitrary set of $J$-factors. This is one of the
most recent in a long line of Fermi-LAT dwarf searches
(e.g.,~\citealt{Fermi-LAT:2015att,Fermi-LAT:2016uux,McDaniel:2023bju}; and the multi-instrument combination of ~\citealt{Fermi-LAT:2025gei}). With their low astrophysical backgrounds, dwarf galaxies
furnish one of the cleanest probes of the thermal relic annihilation cross
section~\citep{Steigman:2012nb}, and they bear directly on the Fermi
Galactic Center Excess (GCE)---an extended $\osim$GeV gamma-ray signal toward the
inner Galaxy
that has been interpreted as flux from annihilating DM~\citep{Murgia:2020dzu}. 

We synthesize the results of our astrophysical $J$-factor modeling into three central conclusions. First, $J$-factors derived from a Jeans analysis are sensitive to prior assumptions. This shifts the resulting annihilation limits by a factor of $\osim$2--4 for the $b\bar{b}$ annihilation channel considered. Second, the highest-$J$-factor dwarfs are spatially extended, challenging the validity of the point-source approximation adopted in many analyses of Fermi data. Third, based on the semi-analytic model and kinematic data, the highest-$J$-factor halos have already been discovered, suggesting future ultra-faint discoveries are unlikely to substantially strengthen limits on annihilating DM. 

This paper is organized as follows. \autoref{sec:methods} provides an overview of
the methods---the semi-analytic \SatGen{} modeling and the Jeans analyses---along
with the observational data used in this work.  \autoref{sec:jfactor} presents the $J$-factor and angular extent results for the different dwarf galaxies and discusses their implications for annihilation cross-section limits. 
\autoref{sec:discussion} explores the prospects of improving constraints with future dwarf discoveries and discusses the implication of these results for the GCE. \autoref{sec:conclusions} concludes. 
\autoref{app:jeans} overviews the Jeans analysis procedure. 
\autoref{app:all_priors} discusses the effects of a range of informative priors on the Jeans analysis results. 
\autoref{app:craterantlia} discusses the inconsistency of the Jeans analysis results with the inference methods presented in \paperone{} for two dwarf galaxies, Crater~II and Antlia~II.
\autoref{app:limit_comparison} compares the upper limits on DM annihilation presented in this work to previous results from the literature. 

\Needspace{4\baselineskip}
\section{Methodology}
\label{sec:methods}

The astrophysical $J$-factor relevant for DM indirect detection is defined as 
\es{eq:J}{
J \equiv \int d \Omega \int ds\, \rho^2_{\scriptscriptstyle{\rm DM}}(s,\Omega) \,,
}
where $\rho_{\scriptscriptstyle{\rm DM}}$ is the DM density, $s$ is the LOS distance from Earth, and the integral is taken over solid angle $\Omega$~\citep[see, e.g.,][]{Safdi:2022xkm}.\footnote{The $J$-factor is sometimes defined without the solid-angle integral, especially when discussing the extended DM halo of the MW~\citep{Safdi:2022xkm}.} Focusing specifically on dwarf galaxies, we adopt a $0.5^\circ$ radius around each target as our benchmark region of interest~(ROI), following the conventions of \citet{Pace:2018tin,Circiello:2026inp}.  We calculate the dwarf $J$-factors using two distinct approaches, each with unique advantages. The first approach probabilistically infers values using satellites generated with \SatGen{} (\autoref{sec:satgen}), while the second relies on a standard Jeans analysis (\autoref{sec:jeans}).

\Needspace{4\baselineskip}
\subsection{\SatGen{} Inference}
\label{sec:satgen}

This subsection briefly reviews the inference procedure used in \paperone{} to infer the densities of MW satellites from those generated semi-analytically with \SatGen{}~\citep{Jiang:2020rdj,Green:2021vkf}.\footnote{\href{https://github.com/JiangFangzhou/SatGen}{https://github.com/JiangFangzhou/SatGen}}  Here, we extend the procedure to infer satellite $J$-factors. \SatGen{} can produce a large population of subhalos
consistent with the hierarchical assembly of a MW--mass~($10^{12}\,\Msun{}$)
host. It builds halo merger trees, seeds each progenitor with an NFW profile
and a cosmologically motivated concentration, and integrates the subhalo orbit
in a time-evolving host potential while tracking tidal mass loss. We use the Fiducial \SatGen{} run of \paperone{}, which resolves halos down to $M_{\rm peak} = 10^8\,\Msun{}$. Code used to produce our \SatGen{} runs can be found in our fork of the \SatGen{} GitHub repository.\footnote{\href{https://github.com/folsomde/SatGen}{https://github.com/folsomde/SatGen}}

Following
\citet{Folsom:2023ejk}, we infer an unobserved halo property $\mathbf{X}$ of a given
satellite by conditioning this population on an observable $\mathbf{\Theta}$,
\begin{equation}
\label{eq:inference}
f(\mathbf{X}) \approx \int f_{\rm pred}(\mathbf{X},\mathbf{\Theta})\,
\frac{f(\mathbf{\Theta})}{f_{\rm pred}(\mathbf{\Theta})}\,\mathrm{d}\mathbf{\Theta}\,,
\end{equation}
where $f_{\rm pred}$ is the distribution of \SatGen{} satellites and $f(\mathbf{\Theta})$ is
the observed probability distribution function~(PDF) of $\mathbf{\Theta}$; in practice, $f(\mathbf{X})$ is a weighted sum over
\SatGen{} halos, each carrying a weight $f(\mathbf{\Theta})/f_{\rm pred}(\mathbf{\Theta})$. For each MW dwarf, we restrict to \SatGen{} satellites within
$50$--$200\%$ of its measured Galactocentric distance.  We then infer a $J$-factor PDF by placing each \SatGen{} satellite
that passes this selection cut at the distance of the observed dwarf from Earth,
so that $\mathbf{X}$ is the $J$-factor as viewed from that
distance.

As in \paperone{}, we consider two conditioning observables. The
\Mhalf{} inference takes $\mathbf{\Theta} = \Mhalf{}$, the dynamical mass enclosed within the
observed 3D half-light radius, $\rhalf$.  This is obtained from the LOS velocity dispersion, $\sigmaLOS{}$, via the
\citealt{Wolf:2009tu} estimator, 
\begin{equation}
    \Mhalf \approx 3 G^{-1} \sigmaLOS^2 \rhalf \,,
\end{equation}
where $G$ is the gravitational constant. The \Mstar{} inference instead takes
$\mathbf{\Theta} = \Mstar{}$, the observed stellar mass, with \SatGen{} halos assigned stellar
masses at infall through a SHMR. We adopt the \Fattahi{} relation for our primary comparison, but also explore other variations. 

\begin{figure}[t]
    \includegraphics[width=\columnwidth]{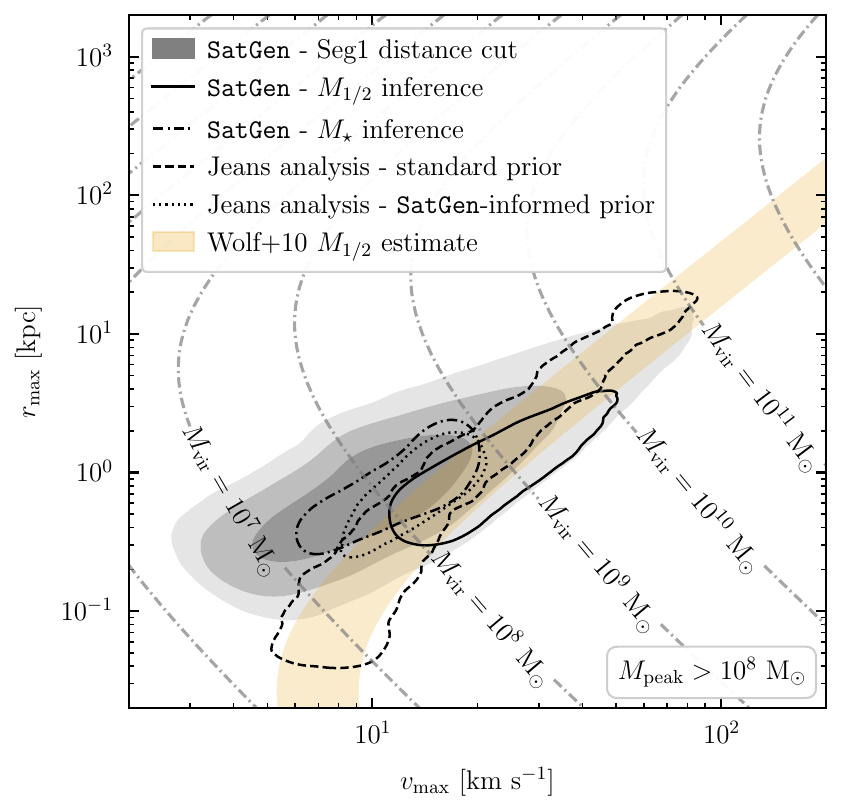}
    \caption{Inferred values for the $z=0$ maximum circular velocity \vmax{} and the radius \rmax{} at which it occurs for the MW dwarf Segue~1~(Seg1). In gray are the 68, 95, and 99.5\% containment regions for the \SatGen{} satellite halos satisfying the distance selection for Seg1.
    The solid and dash-dotted black contours show the 68\% containment for the profile parameters inferred using \Mhalf{} and \Mstar{}, respectively. The \Mstar{} inference assumes the SHMR from \Fattahi{}. Both the \Mhalf{}- and \Mstar{}-inference results are computed using the Fiducial \SatGen{} run and Seg1 distance selection.
   The dashed and dotted black lines correspond to the results for the Jeans analyses with the standard prior and \SatGen{}-informed prior, respectively.
    The light orange band shows the locus of NFW halo parameters whose enclosed mass at the half-light radius of Seg1 falls within the 68\% confidence interval of the observed $\Mhalf{}$ (as computed using the \Wolf{} estimator), which the \Mhalf{} inference recovers.
    The gray dot-dashed contours denote lines of constant \Mvir{}, the virial mass of a NFW halo with the given parameters at $z=0$.
    }
    \label{fig:rmaxvmax}
\end{figure}

To illustrate how the choice of conditioning observable affects the recovered properties of a dwarf's density distribution, \autoref{fig:rmaxvmax} presents the results for the specific case of Segue~1.  The gray-shaded regions show the 68, 95, and 99.5\% containment of $\vmax$ and $\rmax$ for all \SatGen{} satellites that pass the galactocentric distance selection for Segue~1. The quantities $\vmax$ and $\rmax$ are the maximum circular velocity within the halo and the corresponding radius at which that circular velocity is reached, respectively. The light orange band indicates the range of parameters consistent within the 68\% confidence interval of the observed $\Mhalf{}$ assuming an NFW profile.  The black solid and dot-dashed lines indicate the 68\% containment for the profile parameters inferred using the $\Mhalf{}$ and $\Mstar{}$ inference, respectively.  Including the kinematic information pushes the inference towards more-concentrated halos, or those with greater virial mass. Analogues of \autoref{fig:rmaxvmax} for other satellites considered in this work can be found in our GitHub repository.\footnote{ \href{https://github.com/kailashraman/SatelliteDensityInference/}{https://github.com/kailashraman/SatelliteDensityInference/}}

Each conditioning observable has distinct strengths and limitations. The $\Mhalf{}$ inference relies on measured LOS velocity dispersions, but is therefore vulnerable to observational biases that could skew results. In contrast, $\Mstar{}$ offers a clearer baseline for cold dark matter~(CDM) expectations via the SHMR, but incorporates no kinematic information. Because the MW is not necessarily a median \SatGen{} host, it is not immediately clear how the $J$-factors inferred in this way generalize to our own Galaxy. In contrast to the Jeans analyses described next, neither the $\Mhalf{}$ nor $\Mstar$ inference make any assumption about equilibrium.

\subsection{Jeans Analyses}
\label{sec:jeans}

To complement the \SatGen{}-inferred $J$-factors, we also perform a Jeans analysis on each dwarf galaxy, relating its halo density profile to its LOS velocity dispersion. Importantly, this procedure assumes each dwarf is spherically symmetric and in steady state. We largely follow \citet{Pace:2018tin}, but use updated kinematic data of individual stars and explore a variety of cosmologically motivated priors.  This subsection highlights relevant details of the method; see \autoref{app:jeans} for a complete overview.

The free parameters of the Jeans analysis include the NFW halo parameters~($r_s$, $\rho_s$), symmetrized anisotropy parameter~($\tilde\beta$), and systemic velocity~($\bar{V}$). We additionally profile over observational uncertainties on the stellar scale radius~($r_p$), heliocentric distance~($d$), and systemic proper motions~($\mu_{\alpha}$, $\mu_{\delta}$). We perform a Bayesian analysis using the Gaussian unbinned log likelihood $\ln \mathcal{L} = \sum_i \ln \mathcal{L}_i$, where the sum is taken over the dwarf's member stars and 
\begin{equation}
  \ln\mathcal{L}_i
  = -\frac{1}{2}\ln \bigl(2\pi s_i^2\bigr)
    - \frac{(V_i - \bar V)^2}{2\,s_i^2} \,,
  \label{eq:per-star-lnlike}
\end{equation}
with $s_i^2 \equiv \sigmaLOS^2(R_i;\,\rho_s, r_s, \beta, r_p)+ \sigma_{\varepsilon,i}^2$~\citep{Strigari:2007at}.  Here, $V_i$ is the stellar LOS velocity, $\sigma_{\varepsilon,i}^2$ is the corresponding measurement uncertainty, and $R_i$ is the distance of the star from the center of the dwarf. Additionally, we correct the LOS velocities that enter the likelihood for perspective motion effects~\citep{Kaplinghat:2008sm,Walker:2008ji}.
Posterior sampling is conducted via the nested-sampling Python package \texttt{dynesty}~\citep{Speagle:2019ivv,sergey_koposov_2025_17268284}, which implements the algorithm of \citet{Skilling:2004pqw,Skilling:2006gxv} with multi-ellipsoidal bounds~\citep{Feroz:2008xx} and the random walk evolution scheme~\citep{Skilling:2006gxv}.\footnote{Code for reproducing our Jeans analysis results is located at \href{https://github.com/kailashraman/DwarfJeansAnalysis/}{https://github.com/kailashraman/DwarfJeansAnalysis/}.} 

The choice of priors for this Bayesian analysis can affect the recovered $J$-factors~\citep[see, e.g.,][]{Ando:2020yyk,Horigome:2022gge}. The standard
procedure is to adopt wide, uninformative priors on the halo structural parameters
so that the kinematic data dominate the posterior wherever they are constraining.
For dwarf galaxies with few member stars, however, the likelihood is too weak to
determine the halo parameters, and the posterior is instead governed by the prior; an
unrestricted prior can then favor halo parameters that are unphysical and not realized in the CDM population. \SatGen{} offers a natural remedy: because it predicts
the cosmological distribution of halo parameters, we can use it to build
physics-informed priors that recover a sharply constrained, data-dominated posterior
where the kinematics are informative while defaulting to the CDM expectation where
they are not. We fold progressively more \SatGen{} information into the prior,
yielding the following sequence of choices, ordered from least to most informative:
\begin{itemize}
    \item A wide log-uniform prior, almost identical to that of~\citet{Pace:2018tin}:
    log-uniform on $-2<\log_{10} r_s/{\rm kpc}<1$ and
    $4<\log_{10} \rho_s/{{\rm \Msun{}~kpc}^{-3}}<14$, with the additional selection $r_s>r_{1/2}$. Unlike ~\citet{Pace:2018tin}, we also impose $\vmax{}>1 \, \kms{}$.\footnote{The extra selection of $\vmax{}>1 \, \kms{}$ is a loose criterion for galaxy formation recommended by the authors of~\citet{Pace:2018tin} in private communication and used in~\citet{DELVE:2024eud, DELVE:2024yct}.}
    This prior---referred to as the \emph{standard prior} throughout this work---lets the kinematic data dominate where they are most constraining.
    \item A log-uniform prior restricted to the two-dimensional region of
    $(r_s,\rho_s)$ spanned by the Fiducial \SatGen{} population (no distance
    selection). Because a \SatGen{} halo is a truncated NFW halo, we approximate it as an NFW profile with matching
    $\rmax{}$ and $\vmax{}$. This prior remains log-flat but excludes halo parameters
    that do not arise in the cosmological population.
    \item A log-normal prior, fit to the \SatGen{} distribution. We perform a log-normal fit to
    the \SatGen{} $r_s$ distribution and a conditional log-normal fit to $\rho_s$ at
    fixed $r_s$ (again the Fiducial run, NFW-approximated, no distance selection).
    Beyond preferring physical regions of parameter space, this additionally encodes the shape of the
    cosmological halo distribution.
    \item A SHMR-weighted log-normal prior.
    We take the \Mstar{}-inferred distribution of $(\rmax{}, \vmax{})$ for the dwarf in question, yielding a SHMR-preferred region of parameter space subject to the dwarf-specific galactocentric distance selection.
    From this distribution, we apply the same log-normal fit procedure described above. 
    The main text presents results for the \Fattahi{}
    relation while other SHMRs are discussed in~\autoref{app:all_priors}.
    This is the most informative prior, folding in the galaxy-formation expectation
    for the dwarf's stellar mass together with its orbital location.  We refer to this as the {\it \SatGen{}-informed prior}.
\end{itemize}
To illustrate the effects of prior choice on the Jeans-inferred densitiy for a given dwarf, \autoref{fig:rmaxvmax} shows the recovered $(\rmax, \vmax)$ for Segue~1 for the standard  and \SatGen{}-informed prior (dashed and dotted black, respectively).  When using the standard prior, the recovered parameters are consistent with those expected for the NFW, but extend to unphysically large virial masses.  Using the \SatGen{}-informed prior leads to results that are more consistent with cosmologically-motivated halo masses. 

Since satellites undergo tidal stripping of their density profiles while orbiting the MW, we must account for such effects to avoid biasing the inferred $J$-factor high. We therefore follow the convention of \citet{Pace:2018tin} and truncate the Jeans analysis-inferred density profiles when computing the $J$-factor at the tidal radius
\begin{equation}
    r_t = d_{\rm GC}\left[ \frac{M_{\rm sub}(<r_t)}{\left(2 - \frac{d \ln M_{\rm host}}{d \ln r}\right) M_{\rm host}(d_{\rm GC})} \right]^{1/3} \, ,
\end{equation}
where $M_{\rm sub}(<r_t)$ is the subhalo's enclosed mass within its tidal radius, and $M_{\rm host}(d_{\rm GC})$ is the host halo enclosed mass at galactocentric radius $d_{\rm GC}$~\citep{Springel:2008cc}.\footnote{Similar to the dwarf galaxy specific distance selection, we make the approximation of using current satellite galactocentric distances instead of satellite pericenters.} For all of the dwarf galaxies, with the exception of Crater~II, Antlia~II, and Bo\"otes~III, the fraction of the stellar profile beyond the 95th percentile of $r_t$ is less than 5\%. Since these dwarf galaxies are subdominant for DM annihilation searches, we make the approximation of truncating the profiles after performing the Jeans analyses. For the host halo mass model, we adopt the \texttt{MilkyWayPotential2022} model
from \texttt{Gala}~\citep{gala,adrian_price_whelan_2025_16923466}, with a disk calibrated to the
rotation curve of \citet{Eilers:2019gqs} and the vertical phase-space structure of
\citet{darraghford2023textttescargotmappingverticalphase}. We use this host halo mass model since it is a snapshot of the current MW, constrained by enclosed-mass measurements.

\subsection{Data}
\label{sec:data}

We study the same sample of 39 MW dwarf galaxies as \paperone{}. 
As in that work, we divide the sample by stellar mass and the number of resolved member stars into the three categories: $\Mstar{} > 10^5~\Msun{}$, Ultra-faints, and systems with fewer than 10 stars with spectroscopic LOS velocity measurements ($N_{\rm stars} < 10$).
Each dwarf's
structural and systemic parameters---half-light radius, ellipticity, proper motion,
sky position, and heliocentric distance---are taken from the Local Volume Database
\citep[LVDB;][]{Pace:2024sys}. The Jeans analyses further require the LOS velocities of individual member stars. For these, we use per-star measurements from
the Keck/Deimos Stellar Archive~\citep[KDSA;][]{geha2026keckdeimosstellararchivei,geha2026keckdeimosstellararchiveii}
where available, and otherwise from the original spectroscopic studies underlying the
corresponding LVDB velocity dispersions; the per-dwarf data sources are listed in
\paperone{}.

Following \citet{geha2026keckdeimosstellararchiveii}, we
combine repeat measurements for stars observed over multiple epochs, flag candidate
binaries through a velocity-variability test, and retain only member stars with
membership probability above $0.5$ that lie within twice the projected half-light
radius. The complete procedure, including our treatment of multi-instrument systematic
offsets, is detailed in \autoref{app:jeans}. Critically, these are the same stellar
data that set the velocity dispersions used in the \Mhalf{} inference, so the \Mhalf{} and Jeans inferences can be compared directly at the
level of the input data.

\begin{deluxetable*}{lcccccccccc}[!tp]
\tablecaption{Measured and inferred properties of the dwarf galaxies considered in this work.}
\tablewidth{0pt}
\tabletypesize{\small}
\setlength{\tabcolsep}{2pt}
\renewcommand{\arraystretch}{1.0}
\tablehead{
\colhead{Name} & \colhead{$d$} & \colhead{\rhalf{}} & \colhead{\sigmaLOS{}} & \colhead{$\log_{10}\frac{L_V}{\Lsun{}}$} & \colhead{$\log_{10}\frac{\Mhalf{}}{\Msun{}}$} & \colhead{$\log_{10}\frac{\Mstar{}}{\Msun{}}$} & \colhead{$\log_{10}\frac{J_{\Mhalf{}}}{\Junits{}}$} & \colhead{$\log_{10}\frac{J_{\rm Jeans,stand.}}{\Junits{}}$} & \colhead{$\log_{10}\frac{J_{\rm Jeans,\SatGen{}}}{\Junits{}}$} \\
\colhead{} & \colhead{[kpc]} & \colhead{[pc]} & \colhead{[\kms{}]} & \colhead{} & \colhead{} & \colhead{} & \colhead{} & \colhead{} & \colhead{}
}
\startdata
\multicolumn{10}{c}{$\Mstar{}>10^5~\Msun{}$} \\
\hline
AntII$^b$ & $124^{+5}_{-5}$ & $3200^{+300}_{-300}$ & $6.0^{+0.4}_{-0.4}$ & $5.82^{+0.07}_{-0.07}$ & $7.90^{+0.12}_{-0.12}$ & $5.90^{+0.23}_{-0.23}$ & $17.5^{+0.3}_{-0.3}$ & $16.85^{+0.14}_{-0.12}$ & $16.99^{+0.11}_{-0.10}$ \\
CVnI$^a$ & $211^{+6}_{-6}$ & $436^{+21}_{-21}$ & $7.7^{+0.4}_{-0.4}$ & $5.42^{+0.05}_{-0.05}$ & $7.25^{+0.11}_{-0.11}$ & $5.50^{+0.21}_{-0.21}$ & $17.42^{+0.13}_{-0.12}$ & $17.45^{+0.14}_{-0.10}$ & $17.48^{+0.11}_{-0.11}$ \\
Car$^b$ & $106^{+5}_{-5}$ & $331^{+17}_{-17}$ & $6.6^{+1.2}_{-1.2}$ & $5.70^{+0.06}_{-0.06}$ & $7.00^{+0.19}_{-0.19}$ & $5.78^{+0.22}_{-0.22}$ & $17.8^{+0.3}_{-0.3}$ & $17.88^{+0.10}_{-0.10}$ & $17.88^{+0.10}_{-0.08}$ \\
CraII$^b$ & $117^{+4}_{-4}$ & $1280^{+130}_{-120}$ & $2.3^{+0.4}_{-0.3}$ & $5.21^{+0.07}_{-0.07}$ & $6.69^{+0.19}_{-0.16}$ & $5.28^{+0.23}_{-0.23}$ & $16.6^{+0.4}_{-0.4}$ & $15.18^{+0.24}_{-0.30}$ & $16.34^{+0.19}_{-0.21}$ \\
Dra$^a$ & $81.5^{+1.5}_{-1.5}$ & $258^{+6}_{-6}$ & $9.55^{+0.26}_{-0.25}$ & $5.48^{+0.04}_{-0.04}$ & $7.21^{+0.10}_{-0.10}$ & $5.56^{+0.20}_{-0.20}$ & $18.85^{+0.10}_{-0.07}$ & $18.90^{+0.07}_{-0.08}$ & $18.83^{+0.05}_{-0.06}$ \\
LeoI$^a$ & $258^{+10}_{-10}$ & $306^{+25}_{-25}$ & $9.29^{+0.26}_{-0.26}$ & $6.66^{+0.14}_{-0.14}$ & $7.26^{+0.11}_{-0.11}$ & $6.7^{+0.3}_{-0.3}$ & $17.76^{+0.16}_{-0.11}$ & $17.75^{+0.10}_{-0.10}$ & $17.74^{+0.09}_{-0.09}$ \\
LeoII$^a$ & $233^{+14}_{-14}$ & $220^{+13}_{-13}$ & $7.5^{+0.4}_{-0.4}$ & $5.82^{+0.07}_{-0.07}$ & $6.94^{+0.11}_{-0.11}$ & $5.90^{+0.23}_{-0.23}$ & $17.64^{+0.19}_{-0.13}$ & $17.70^{+0.20}_{-0.16}$ & $17.68^{+0.14}_{-0.12}$ \\
Scl$^a$ & $83.9^{+1.5}_{-1.5}$ & $298^{+6}_{-6}$ & $8.7^{+0.3}_{-0.3}$ & $6.24^{+0.07}_{-0.07}$ & $7.19^{+0.11}_{-0.11}$ & $6.32^{+0.23}_{-0.23}$ & $18.57^{+0.09}_{-0.08}$ & $18.58^{+0.11}_{-0.09}$ & $18.56^{+0.08}_{-0.08}$ \\
Sex$^a$ & $86^{+4}_{-4}$ & $650^{+40}_{-40}$ & $8.8^{+0.5}_{-0.5}$ & $5.42^{+0.06}_{-0.06}$ & $7.54^{+0.11}_{-0.11}$ & $5.50^{+0.22}_{-0.22}$ & $18.22^{+0.16}_{-0.13}$ & $18.11^{+0.18}_{-0.18}$ & $18.22^{+0.15}_{-0.14}$ \\
UMi$^a$ & $70^{+4}_{-4}$ & $334^{+17}_{-17}$ & $8.86^{+0.27}_{-0.24}$ & $5.48^{+0.06}_{-0.06}$ & $7.26^{+0.11}_{-0.11}$ & $5.56^{+0.22}_{-0.22}$ & $18.69^{+0.08}_{-0.07}$ & $18.78^{+0.10}_{-0.11}$ & $18.72^{+0.11}_{-0.10}$ \\
\hline
\multicolumn{10}{c}{Ultra-faints} \\
\hline
Bo\"oI$^a$ & $66.4^{+2.4}_{-2.4}$ & $215^{+11}_{-11}$ & $3.2^{+0.5}_{-0.4}$ & $4.34^{+0.13}_{-0.13}$ & $6.18^{+0.17}_{-0.15}$ & $4.42^{+0.29}_{-0.29}$ & $17.18^{+0.27}_{-0.26}$ & $17.15^{+0.29}_{-0.28}$ & $17.38^{+0.24}_{-0.21}$ \\
Bo\"oII$^a$ & $41.7^{+1.2}_{-1.2}$ & $44^{+7}_{-7}$ & $1.9^{+0.8}_{-0.6}$ & $3.11^{+0.12}_{-0.12}$ & $5.1^{+0.4}_{-0.3}$ & $3.19^{+0.28}_{-0.28}$ & $17.9^{+0.7}_{-0.6}$ & $17.4^{+0.7}_{-0.8}$ & $18.3^{+0.4}_{-0.4}$ \\
Bo\"oIII$^a$ & $46.6^{+0.4}_{-0.4}$ & $600^{+70}_{-70}$ & $5.3^{+2.1}_{-1.7}$ & $4.23^{+0.21}_{-0.21}$ & $7.1^{+0.4}_{-0.3}$ & $4.3^{+0.4}_{-0.4}$ & $18.0^{+0.6}_{-0.5}$ & $17.9^{+0.6}_{-0.6}$ & $18.6^{+0.3}_{-0.3}$ \\
CVnII$^a$ & $160^{+4}_{-4}$ & $73^{+14}_{-14}$ & $5.3^{+1.3}_{-1.0}$ & $4.00^{+0.15}_{-0.15}$ & $6.15^{+0.24}_{-0.21}$ & $4.1^{+0.3}_{-0.3}$ & $18.1^{+0.6}_{-0.4}$ & $18.2^{+0.4}_{-0.5}$ & $17.89^{+0.24}_{-0.25}$ \\
CarII$^b$ & $37.4^{+0.4}_{-0.4}$ & $102^{+10}_{-10}$ & $3.4^{+1.2}_{-0.8}$ & $3.76^{+0.05}_{-0.05}$ & $5.9^{+0.3}_{-0.2}$ & $3.84^{+0.21}_{-0.21}$ & $18.3^{+0.6}_{-0.5}$ & $18.2^{+0.6}_{-0.6}$ & $18.6^{+0.3}_{-0.4}$ \\
CenI$^b$ & $118^{+4}_{-4}$ & $95^{+12}_{-12}$ & $4.2^{+0.6}_{-0.5}$ & $4.07^{+0.10}_{-0.10}$ & $6.07^{+0.17}_{-0.15}$ & $4.15^{+0.26}_{-0.26}$ & $17.7^{+0.3}_{-0.3}$ & $17.8^{+0.4}_{-0.3}$ & $17.88^{+0.22}_{-0.23}$ \\
CB$^a$ & $42.3^{+1.6}_{-1.6}$ & $73^{+6}_{-6}$ & $3.3^{+0.6}_{-0.5}$ & $3.65^{+0.13}_{-0.13}$ & $5.75^{+0.18}_{-0.16}$ & $3.73^{+0.29}_{-0.29}$ & $18.3^{+0.3}_{-0.3}$ & $18.5^{+0.3}_{-0.4}$ & $18.47^{+0.23}_{-0.26}$ \\
EriII$^b$ & $370^{+9}_{-9}$ & $239^{+16}_{-16}$ & $6.9^{+1.2}_{-0.9}$ & $4.78^{+0.14}_{-0.14}$ & $6.90^{+0.18}_{-0.15}$ & $4.86^{+0.30}_{-0.30}$ & $17.1^{+0.3}_{-0.3}$ & $17.3^{+0.4}_{-0.3}$ & $17.20^{+0.17}_{-0.19}$ \\
EriIV$^a$ & $70^{+4}_{-4}$ & $75^{+13}_{-13}$ & $4.5^{+1.1}_{-0.9}$ & $3.35^{+0.14}_{-0.14}$ & $6.04^{+0.25}_{-0.21}$ & $3.43^{+0.30}_{-0.30}$ & $18.5^{+0.5}_{-0.4}$ & $18.5^{+0.4}_{-0.5}$ & $18.31^{+0.28}_{-0.27}$ \\
Her$^a$ & $131^{+6}_{-6}$ & $159^{+17}_{-17}$ & $2.2^{+0.7}_{-0.6}$ & $4.25^{+0.11}_{-0.11}$ & $5.75^{+0.28}_{-0.26}$ & $4.33^{+0.27}_{-0.27}$ & $16.2^{+0.6}_{-0.5}$ & $16.3^{+0.6}_{-0.6}$ & $17.0^{+0.4}_{-0.4}$ \\
HyiI$^b$ & $27.5^{+0.5}_{-0.5}$ & $70^{+9}_{-6}$ & $2.7^{+0.5}_{-0.4}$ & $3.82^{+0.05}_{-0.05}$ & $5.55^{+0.20}_{-0.17}$ & $3.90^{+0.21}_{-0.21}$ & $18.3^{+0.3}_{-0.3}$ & $18.5^{+0.3}_{-0.3}$ & $18.6^{+0.3}_{-0.3}$ \\
LeoIV$^a$ & $151^{+4}_{-4}$ & $136^{+17}_{-17}$ & $3.3^{+0.9}_{-0.7}$ & $3.91^{+0.13}_{-0.13}$ & $6.01^{+0.26}_{-0.21}$ & $3.99^{+0.29}_{-0.29}$ & $16.9^{+0.4}_{-0.4}$ & $16.8^{+0.4}_{-0.4}$ & $17.4^{+0.3}_{-0.5}$ \\
LeoV$^a$ & $169^{+5}_{-5}$ & $49^{+19}_{-19}$ & $3.4^{+1.7}_{-1.0}$ & $3.69^{+0.17}_{-0.17}$ & $5.6^{+0.5}_{-0.3}$ & $3.8^{+0.3}_{-0.3}$ & $17.7^{+1.0}_{-0.7}$ & $17.2^{+0.8}_{-0.8}$ & $17.6^{+0.3}_{-0.3}$ \\
PegIII$^a$ & $215^{+12}_{-12}$ & $110^{+20}_{-17}$ & $2.8^{+1.2}_{-0.9}$ & $3.60^{+0.14}_{-0.12}$ & $5.8^{+0.4}_{-0.3}$ & $3.68^{+0.30}_{-0.28}$ & $16.5^{+0.7}_{-0.7}$ & $15.9^{+0.8}_{-1.4}$ & $17.2^{+0.3}_{-0.5}$ \\
PegIV$^a$ & $89.9^{+1.2}_{-1.2}$ & $56^{+12}_{-9}$ & $3.3^{+1.2}_{-0.9}$ & $3.63^{+0.09}_{-0.09}$ & $5.6^{+0.3}_{-0.3}$ & $3.71^{+0.25}_{-0.25}$ & $18.0^{+0.6}_{-0.5}$ & $17.7^{+0.7}_{-0.6}$ & $18.0^{+0.3}_{-0.4}$ \\
RetII$^b$ & $31.6^{+1.5}_{-1.5}$ & $49^{+7}_{-7}$ & $3.6^{+1.0}_{-0.7}$ & $3.17^{+0.08}_{-0.08}$ & $5.65^{+0.27}_{-0.21}$ & $3.25^{+0.24}_{-0.24}$ & $19.0^{+0.5}_{-0.4}$ & $18.9^{+0.4}_{-0.4}$ & $18.85^{+0.29}_{-0.28}$ \\
Seg1$^a$ & $22.9^{+2.1}_{-2.1}$ & $26^{+4}_{-4}$ & $4.0^{+1.0}_{-0.9}$ & $2.5^{+0.4}_{-0.4}$ & $5.46^{+0.25}_{-0.22}$ & $2.5^{+0.5}_{-0.5}$ & $19.9^{+0.5}_{-0.4}$ & $19.5^{+0.4}_{-0.5}$ & $19.3^{+0.3}_{-0.3}$ \\
TucIV$^b$ & $47^{+4}_{-4}$ & $132^{+24}_{-20}$ & $4.3^{+1.7}_{-1.0}$ & $3.13^{+0.23}_{-0.19}$ & $6.2^{+0.4}_{-0.2}$ & $3.2^{+0.4}_{-0.4}$ & $18.4^{+0.6}_{-0.5}$ & $18.7^{+0.7}_{-0.5}$ & $18.6^{+0.3}_{-0.3}$ \\
UMaI$^a$ & $97^{+6}_{-6}$ & $201^{+16}_{-16}$ & $7.2^{+1.2}_{-1.0}$ & $3.99^{+0.20}_{-0.20}$ & $6.86^{+0.18}_{-0.16}$ & $4.1^{+0.4}_{-0.4}$ & $18.4^{+0.3}_{-0.3}$ & $18.35^{+0.27}_{-0.29}$ & $18.23^{+0.22}_{-0.21}$ \\
UMaII$^a$ & $34.7^{+2.1}_{-2.1}$ & $123^{+10}_{-10}$ & $6.3^{+1.1}_{-0.9}$ & $3.70^{+0.16}_{-0.16}$ & $6.53^{+0.18}_{-0.17}$ & $3.8^{+0.3}_{-0.3}$ & $19.2^{+0.3}_{-0.3}$ & $19.2^{+0.3}_{-0.3}$ & $19.07^{+0.22}_{-0.23}$ \\
Wil1$^a$ & $38^{+7}_{-7}$ & $27^{+6}_{-6}$ & $4.0^{+0.9}_{-0.7}$ & $2.9^{+0.5}_{-0.5}$ & $5.47^{+0.24}_{-0.21}$ & $3.0^{+0.6}_{-0.6}$ & $19.5^{+0.5}_{-0.4}$ & $19.2^{+0.6}_{-0.5}$ & $18.9^{+0.3}_{-0.4}$ \\
\hline
\multicolumn{10}{c}{$N_{\rm stars}<10$} \\
\hline
AqrII$^b$ & $108^{+3}_{-3}$ & $166^{+29}_{-29}$ & $4.7^{+1.8}_{-1.2}$ & $3.68^{+0.08}_{-0.08}$ & $6.4^{+0.4}_{-0.3}$ & $3.76^{+0.24}_{-0.24}$ & $17.7^{+0.7}_{-0.5}$ & $17.9^{+0.6}_{-0.6}$ & $18.0^{+0.3}_{-0.4}$ \\
CarIII$^b$ & $27.8^{+1.3}_{-1.3}$ & $27^{+9}_{-9}$ & $6^{+4}_{-2}$ & $2.89^{+0.12}_{-0.12}$ & $5.8^{+0.7}_{-0.4}$ & $2.97^{+0.28}_{-0.28}$ & $20.3^{+0.8}_{-0.8}$ & $20.3^{+0.9}_{-1.0}$ & $19.2^{+0.4}_{-0.4}$ \\
GruI$^b$ & $126^{+6}_{-6}$ & $150^{+20}_{-30}$ & $2.5^{+1.3}_{-0.8}$ & $3.58^{+0.16}_{-0.16}$ & $5.8^{+0.5}_{-0.3}$ & $3.7^{+0.3}_{-0.3}$ & $16.6^{+0.8}_{-0.8}$ & $16.2^{+0.7}_{-0.7}$ & $17.6^{+0.4}_{-0.5}$ \\
HorI$^b$ & $79^{+4}_{-4}$ & $42^{+4}_{-4}$ & $4.9^{+2.8}_{-0.9}$ & $3.28^{+0.08}_{-0.06}$ & $5.9^{+0.5}_{-0.2}$ & $3.36^{+0.24}_{-0.22}$ & $19.4^{+0.9}_{-0.7}$ & $19.5^{+0.8}_{-0.8}$ & $18.5^{+0.3}_{-0.3}$ \\
LeoVI$^b$ & $111^{+4}_{-6}$ & $120^{+40}_{-60}$ & $2.9^{+1.6}_{-1.3}$ & $3.36^{+0.18}_{-0.24}$ & $5.8^{+0.5}_{-0.5}$ & $3.4^{+0.3}_{-0.4}$ & $17.0^{+1.0}_{-1.0}$ & $17.3^{+0.9}_{-0.9}$ & $17.9^{+0.3}_{-0.4}$ \\
PscII$^b$ & $183^{+14}_{-14}$ & $75^{+8}_{-7}$ & $5^{+4}_{-2}$ & $3.64^{+0.13}_{-0.14}$ & $6.2^{+0.6}_{-0.4}$ & $3.7^{+0.3}_{-0.3}$ & $18.2^{+1.2}_{-0.9}$ & $17.8^{+1.1}_{-1.6}$ & $17.6^{+0.3}_{-0.3}$ \\
TucII$^b$ & $56^{+5}_{-5}$ & $220^{+40}_{-50}$ & $3.8^{+1.1}_{-0.7}$ & $3.43^{+0.12}_{-0.12}$ & $6.34^{+0.28}_{-0.22}$ & $3.51^{+0.28}_{-0.28}$ & $17.7^{+0.5}_{-0.4}$ & $17.5^{+0.5}_{-0.5}$ & $18.0^{+0.4}_{-0.4}$ \\
TucV$^b$ & $55^{+5}_{-5}$ & $31^{+10}_{-9}$ & $1.2^{+0.9}_{-0.6}$ & $2.4^{+0.3}_{-0.3}$ & $4.5^{+0.7}_{-0.5}$ & $2.5^{+0.5}_{-0.4}$ & $17.3^{+1.2}_{-1.0}$ & $18.0^{+2.6}_{-2.9}$ & $18.4^{+0.4}_{-0.4}$ \\
\enddata
\tablecomments{Columns: galaxy name; heliocentric distance $d$; projected half-light radius \rhalf{}; line-of-sight velocity dispersion \sigmaLOS{}; $V$-band luminosity $\log_{10}\frac{L_V}{L_\odot}$; half-light dynamical mass $\log_{10}\frac{\Mhalf{}}{M_\odot}$; stellar mass $\log_{10}\frac{\Mstar{}}{M_\odot}$; and astrophysical $J$-factors $\log_{10}\frac{J_{\Mhalf{}}}{\Junits{}}$, $\log_{10}\frac{J_{\rm Jeans,stand.}}{\Junits{}}$, $\log_{10}\frac{J_{\rm Jeans,\SatGen{}}}{\Junits{}}$ from the \Mhalf{} inference and from Jeans analyses with the standard and \SatGen{}-informed priors, respectively. Name superscripts denote velocity dispersions from (a) KDSA~\citep{geha2026keckdeimosstellararchivei} or (b) LVDB~\citep{Pace:2024sys}; all other observables are from the LVDB. Uncertainties use one significant figure (two if the leading digit is 1 or 2), set by the larger error; central values are rounded to match. \label{tab:J_meta}}
\end{deluxetable*}

The full satellite sample and their properties are summarized in \autoref{tab:J_meta}, which  provides their distance from Earth~($d$), $r_{1/2}$, $\sigmaLOS$, the system $V$-band luminosity~($L_V$), $\Mhalf{}$, and $\Mstar{}$. We also provide astrophysical $J$-factors from the \Mhalf{} inference, as well as the Jeans analysis using the standard prior and the \SatGen{}-informed prior.

\Needspace{4\baselineskip}
\section{Implications for Indirect Detection}
\label{sec:jfactor}
 
For a given dwarf, the spread in DM densities inferred using the \Mhalf{} and \Mstar{} inference, as well as the Jeans analyses, translates into a systematic uncertainty on the expected annihilation signal strength. This section provides updated astrophysical $J$-factors and estimates for the angular extent of the signal, as a function of different inference methods. 

\Needspace{4\baselineskip}
\subsection{Astrophysical $J$-factors \& Angular Extant}

\begin{figure*}
    \centering
    \includegraphics[width=0.9\linewidth]{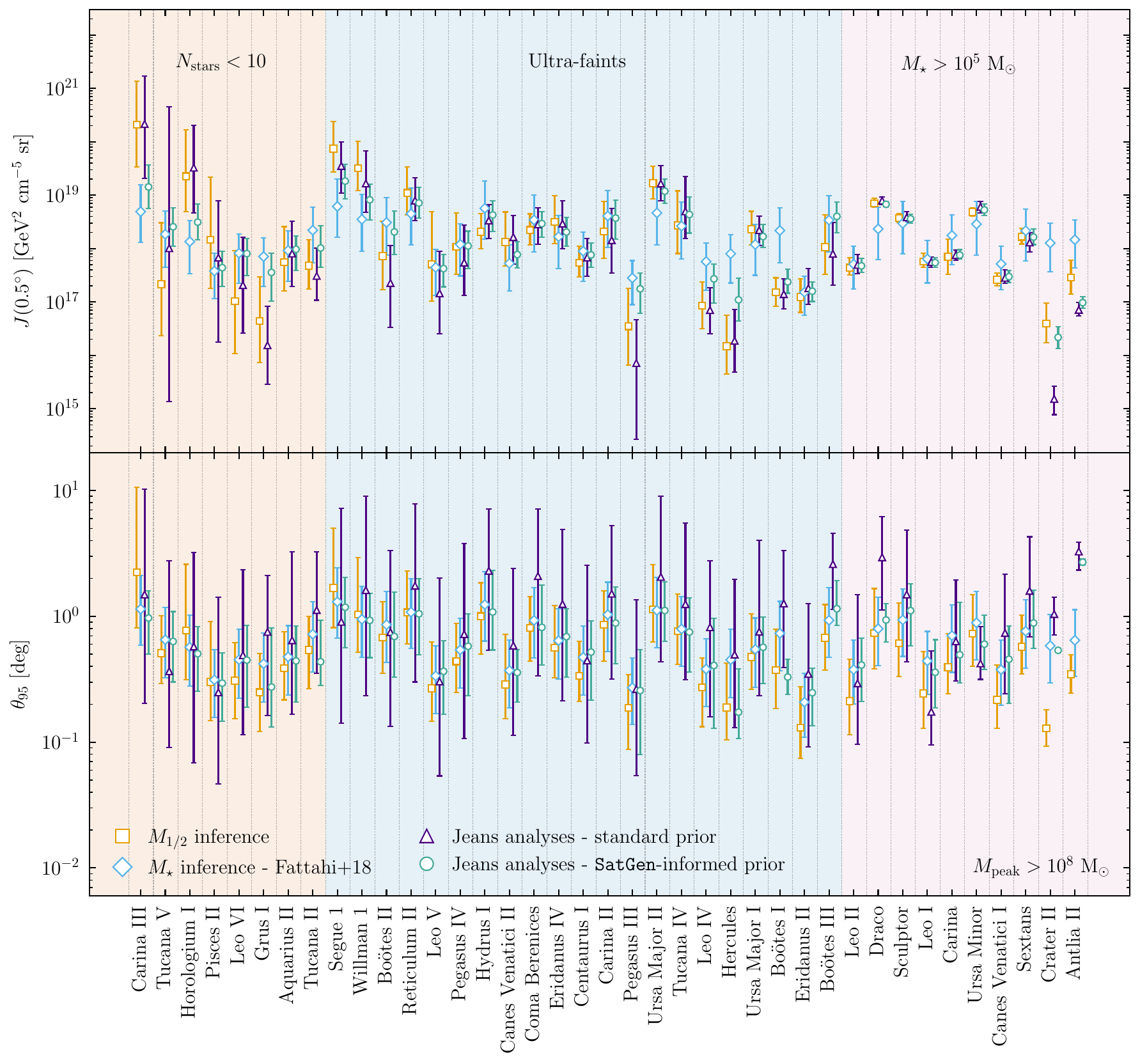}
    \caption{Top: The astrophysical $J$-factor for each MW satellite, integrated to an angular radius of $0.5^{\circ}$. Median results and associated 68\% uncertainties are shown for the \Mhalf{} inference (light orange squares), the \Fattahi{} \Mstar{} inference (light blue diamonds), as well as Jeans analyses using the standard prior~(indigo triangles) and \SatGen{}-informed prior~(teal circles). The MW satellites are ordered by increasing $r_{1/2}$ from left to right within each category ($N_{\rm stars}<10$, Ultra-faints, $\Mstar{} > 10^5~\Msun$).  Compared to using the standard priors, the \SatGen{}-informed priors tend to pull the Jeans analysis results closer to those predicted by the $\Mstar{}$ inference.
    Bottom: The same, but for the 95\% angular containment radius of the astrophysical $J$-factor. Some of the largest $J$-factor dwarfs (e.g., Carina~III, Segue~1, Willman~1) have median 95\% $J$-factor containment larger than $0.5^{\circ}$ across all inference methods.
    }
    \label{fig:Jfactors}
\end{figure*}

The top panel of \autoref{fig:Jfactors} shows the results for the $J$-factor inference for the dwarfs considered in this work. In some cases, the $J$-factors computed through the different inference methods are in reasonable agreement, but there are discrepancies in others.  For the example of Segue~1, there is a $\osim 1\text{--}2 \sigma$ discrepancy between the \Mhalf{}-inferred $J$-factor and that obtained using the \Mstar{} inference. This difference is driven by the fact that, as seen in \autoref{fig:rmaxvmax}, the \Mhalf{} inference picks out denser and/or larger halos.
In general, the $J$-factor differences are correlated with the differences in $\rho_{150}$ (the DM density at 150~pc) and \sigmaLOS{} discussed in \paperone{}. For the same dwarf galaxies as in \paperone{}, we see both large upwards (e.g., Segue~1 and Willman~1) and downwards fluctuations (e.g., Hercules and Bo\"otes~I) of the \Mhalf{} inference results relative to the \Mstar{} inference result.

Appendix A of \paperone{} studies the effects of a variety of systematic changes to the \SatGen{} modeling on the \Mhalf{}-inferred satellite densities, namely $\rho_{150}$. Changes to the concentration-mass relation, dynamical mass estimator, choice of \SatGen{} host mass, and the choice of LMC-associated host selection criteria have a minimal affect on the \Mhalf{}-inferred satellite densities. Additionally, changing the \SatGen{} mass resolution from $\Mpeak{}>10^8\;\Msun{}$ to $\Mpeak{}>10^7\;\Msun{}$ minimally changes the \Mhalf{}-inferred densities. This robustness against modeling choices is driven by the correlation between $\rho_{150}$ and the \sigmaLOS{} data. We have verified that the astrophysical $J$-factor has a similar robustness against \SatGen{} modeling choices.

\autoref{fig:Jfactors} also provides the $J$-factor estimates from the Jeans analyses with the standard prior and the \SatGen{}-informed prior (see \autoref{app:all_priors} for results with the other priors).  Note that, for Tucana~V, only one star survives the Jeans analysis quality cuts and the result is simply determined by the unconstrained log-uniform prior on the halo's structural parameters.

Because the Jeans analyses use the same kinematic datasets as the \Mhalf{}-inference procedure, we compare their $J$-factor predictions. There is good agreement between the \Mhalf{}-inferred $J$-factors and those from the standard-prior Jeans modeling, with a few exceptions. 
The most extreme differences occur for the ultra-diffuse galaxies Crater~II and Antlia~II, with the Jeans analysis predicting $J$-factors 1--2 orders of magnitude below the \Mhalf{} inference. For these dwarf galaxies, the wide log-uniform priors allow the Jeans analysis to pick out NFW halos with $(v_{\rm max}, r_{\rm max})$ values outside of the \SatGen{} distribution, discussed further in \autoref{app:craterantlia}. There is also a large difference for Pegasus~III; the Jeans analysis posterior has a heavy tail at low $J$-factor, pulling the median results $\osim0.7$~dex below the posterior maximum.

Comparing the two Jeans priors, the standard and \SatGen{}-informed results agree
closely for the kinematically well-constrained dwarf galaxies---those with a large
sample of member stars, including most of the $\Mstar{} > 10^5\,\Msun{}$ systems (aside
from Crater~II and Antlia~II)---for which the velocity
data, rather than the prior, set the posterior. The two priors diverge most for the
$N_{\rm stars} < 10$ systems, where the kinematics alone cannot constrain the halo;
there, the \SatGen{}-informed prior pulls the inferred $J$-factor toward the
\Mstar{}-inferred distribution and yields a substantially smaller uncertainty.
Carina~III, for example, has a smaller $J$-factor under the \SatGen{}-informed prior,
as the informative prior draws its large central value downward. The effect is most
striking for Tucana~V, whose single surviving star leaves the $J$-factor essentially
unconstrained under the standard log-uniform prior but yields a well-defined,
physically bounded estimate under the \SatGen{}-informed prior.

\begin{figure*}
\centering
    \includegraphics[width=0.49\textwidth]{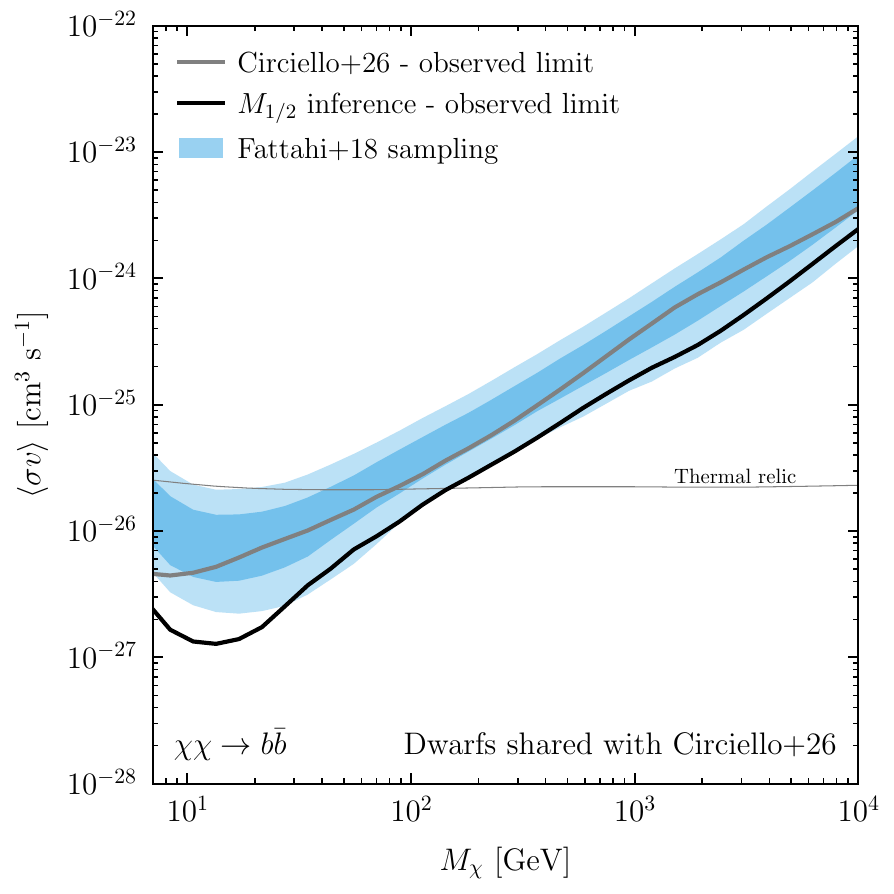}
    \includegraphics[width=0.49\textwidth]{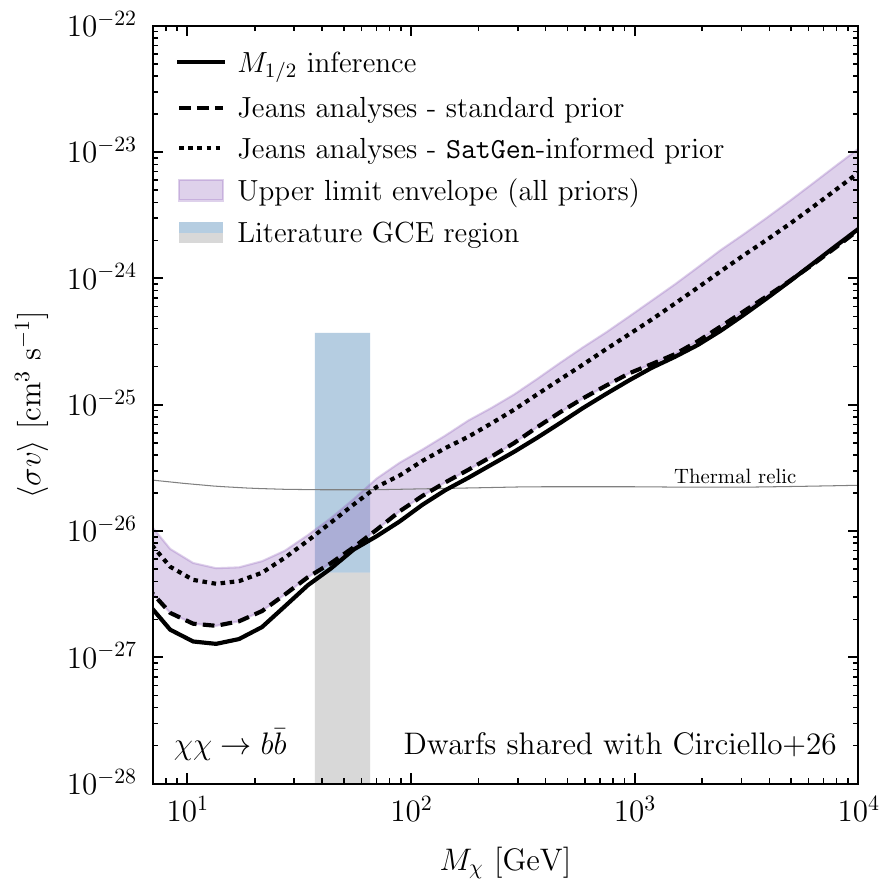}
    \caption{Left: The 95\% confidence upper limit on the $\chi\chi\to b\bar{b}$ velocity-averaged DM annihilation cross section computed by recasting the Fermi-LAT gamma-ray flux limits from~\citet{Circiello:2026inp} for the \Mhalf{}-inferred $J$-factors (solid black). The results are limited to the 30 dwarf galaxies shared between this work and the benchmark sample from~\citet{Circiello:2026inp}. The observed 95\% upper limit from~\citet{Circiello:2026inp}, limited to these shared dwarf galaxies, is shown as the solid gray line. See text for a more detailed comparison of the \citet{Circiello:2026inp} line with the \Mhalf{}-inference line. To compare to cosmological expectations, we show the 68\% and 95\% containment bands for an ensemble of $10^3$ sets of MW dwarf galaxy $J$-factor realizations sampled from the \Fattahi{}-inferred $J$-factor PDFs in light blue, with limits for each $J$-factor set computed using the \Mhalf{} ``measurement" uncertainties. The thermal relic cross section is indicated by the thin solid gray line~\citep{Steigman:2012nb}.
    Right:  For the same 30 dwarf galaxies, the \Mhalf{} inference 95\% upper limit (black, solid), standard-prior Jeans analysis limit (black, dashed), and \SatGen{}-informed-prior Jeans analysis limit (black, dotted). The light purple band corresponds to the envelope of upper limits produced by the 
    Jeans analyses with every prior considered in \autoref{app:jeans}. The cross-section range required to attribute the Galactic Center gamma-ray excess to DM annihilation is shown as a steel blue box (see text), with a gray band extending down to zero to indicate that this excess can be explained in other ways. The use of a \SatGen{}-informed prior can weaken the analysis limits by a factor of $\sim$2--4.
    }
    \label{fig:fermi_reweighting}
\end{figure*}

The bottom panel of \autoref{fig:Jfactors} shows the inference results for $\theta_{\rm 95}$, the angle from the dwarf center for which 95\% of the expected annihilation flux is contained.  Here, we neglect the point-spread function~(PSF) of any individual instrument, though in reality the observed extent of an annihilation signal on the sky is given by the convolution of the true spatial morphology with the instrumental PSF. Strikingly, the Jeans analysis results using the standard prior have much larger uncertainties than the other three cases shown. In particular, some of the highest $J$-factor dwarfs, such as Carina~III, Segue~1, Willman~1, Reticulum~II, and Draco, among others, have $\theta_{95}$ larger than $\osim 0.5^\circ$. Indeed, the median values from the \SatGen{}-informed results are closer to $1^\circ$.  

Many Fermi analyses, such as \citet{McDaniel:2023bju,Circiello:2026inp,Fermi-LAT:2025gei}, treat the dwarf targets as point sources. The 95\% containment of the Fermi PSF for  photons with an incidence angle within $25^{\circ}$ of the Fermi-LAT boresight in the PSF3 event class (the top quartile of Fermi events in terms of angular resolution) ranges from $\osim 1^{\circ}$ at photon energies of 1 GeV to $\osim 0.2^{\circ}$ at photon energy of 10 GeV. Therefore, the point-source assumption is unjustified for the most important targets, except at the lowest DM masses.
\citet{DiMauro:2022hue} find that accounting for source extension can weaken dwarf annihilation limits by up to a factor of $\osim2$, depending on the DM mass. However, previous analyses that include source extension did not include the correlations between $J$-factor and $\theta_{95}$, which would impact the final constraint. Future work should incorporate our inferred halo morphologies in a consistent Fermi data analysis, but this is beyond the scope of the present paper. In the following subsection we simply use the results of \citet{Circiello:2026inp} and rescale by our inferred $J(0.5^\circ)$, but the caveat regarding halo extension should be kept in mind.

\Needspace{4\baselineskip}
\subsection{Dark Matter Annihilation Constraints}
\label{sec:constraints}

We compute joint analysis limits on DM annihilation to $b\Bar{b}$ for DM masses between 7~\GeV{} and 10~\TeV{}
using the data products provided by~\citet{Circiello:2026inp}.\footnote{Data products from~\citet{Circiello:2026inp} can be found  \href{https://figshare.com/articles/dataset/Legacy_analysis_of_Milky_Way_dwarf_spheroidal_satellite_galaxies_an_update/32304966}{here}.} These products allow one to recast the upper limits for different $J$-factors, given log-normal and log-symmetric $J$-factor uncertainties. We approximate the $J$-factor uncertainties for this application by averaging over the asymmetric upper and lower values. The results for this recasting are limited to the set of 30 dwarf galaxies in common between our sample and that of~\citet{Circiello:2026inp}, which includes systems with $N_{\rm stars}<10$.\footnote{Willman~1 is not in the benchmark sample of \citet{Circiello:2026inp} given the literature evidence for tidal disruption.} We again caution that this analysis approximates the halos as point sources, which is not fully justified. 

The solid black line in the left panel of \autoref{fig:fermi_reweighting} shows the 95\% confidence upper limit derived using the $\Mhalf{}$-inferred $J$-factors. For comparison, the solid gray line shows the corresponding limit from \citet{Circiello:2026inp}, limiting to the same set of 30 dwarfs. The $J$-factors used in that work were computed from either a Jeans analysis (using older kinematic datasets than this paper) or the scaling relations of \citet{Pace:2018tin}.
The difference with the \Mhalf{}-inferred limit is driven entirely by the interaction of our updated $J$-factors with various excesses and deficits found in the data. For example, the Tucana~II $J$-factor used by \citet{Circiello:2026inp} is computed using an older kinematic dataset, giving a result that is over an order of magnitude larger than ours. Since the \citet{Circiello:2026inp} data analysis for Tucana~II prefers a positive $\langle \sigma v \rangle$ at low DM masses, using a \emph{smaller} $J$-factor for Tucana~II \emph{strengthens} the limit at these DM masses, creating a ``dip" in the \Mhalf{} result at $M_{\chi}\sim20~\GeV{}$. 

Separately, the \Mhalf{}-inferred $J$-factor for Segue~1 is 0.8~dex larger than the \citet{Pace:2018tin} result quoted by \citet{Circiello:2026inp}, which used an older kinematic dataset and different stellar membership selections. This leads to a strengthening of the \Mhalf{}-inferred limit relative to the \citet{Circiello:2026inp} result by a factor of 2.5 at $M_{\chi}=10$~GeV (deepening the ``dip'') and 1.25--1.75 at $M_{\chi}>30$~GeV. The \Mhalf{}-inferred $J$-factors for Carina~III, 0.6~dex larger than the \citet{Circiello:2026inp} $J$-factor, also modifies the upper limit by up to a factor of two in certain mass ranges. Depending on the energies at which each dwarf galaxy has excesses and deficits in gamma-ray flux, the same $J$-factor change can both strengthen the limit at some masses and weaken the limit at others.

To put the $\Mhalf{}$ expectation in a cosmological context, we derive the expected sensitivity assuming a specific SHMR relation.  In particular, we consider a galaxy's \Mstar{}-inferred $J$-factor PDF to represent a simulation-informed model for the distribution from which its true $J$-factor is sampled. We then sample from these PDFs to create a mock dataset of dwarf galaxy $J$-factors. The uncertainty on this sampled $\log_{10}J$ is set equal to that inferred from the \Mhalf{} analysis for each dwarf, mimicking the measurement uncertainty induced from observing the dwarf's stellar kinematics. Repeating this process $10^3$ times for the \Fattahi{} SHMR yields the light-blue bands in \autoref{fig:fermi_reweighting}~(left), which encompass 68\% and 95\% of the expected limits. Notably, the $\Mhalf{}$-inferred limit lies near the lower edge of the 95\% containment interval of the $\Mstar$-inferred limits. This suggests that the MW is different from a typical $\Mstar$-sampled \SatGen{} realization because it has nearby satellites that are denser than expected. It is also possible that there are significant systematic effects that lead to larger $M_{1/2}$ inferences from the stellar kinematics so that the MW is not as much of an outlier with regard to having denser satellites. 

The right panel of \autoref{fig:fermi_reweighting} focuses on how the priors used in the Jeans analysis affect the recovered limits.  To ground the comparison, the solid black line reproduces the \Mhalf{}-inferred limit provided in the left panel.  The dashed black line, which corresponds to the Jeans analysis result using the standard prior, is in close agreement
with the \Mhalf{} result.  This is unsurprising given that they take the same kinematic data as inputs.

The shaded purple region in the right panel of \autoref{fig:fermi_reweighting} traces the envelope of upper limits using $J$-factor sets from Jeans analyses with every prior considered in this work (see \autoref{sec:jeans}).  The \SatGen{}-informed prior, corresponding to the dotted black line, sets one of the weakest constraints because reweighting the posteriors  using the \Fattahi{} SHMR lowers the inferred densities---and hence the $J$-factors---of the nearby,
high-$J$ dwarfs that dominate the bound. The total width of the purple band is about a factor of 3 at $M_{\chi}=10\;{\rm GeV}$, a factor of 2--2.5 from $M_{\chi}=20\;{\rm GeV}$ to $M_{\chi}=1000\;{\rm GeV}$ and a factor of 3--4 at masses above 1000~\GeV{}. 
This is a direct measure of the
prior dependence of the annihilation limit. 

For the interested reader, \autoref{app:limit_comparison} compares the recasted upper limits from this section to previous literature results.

\Needspace{4\baselineskip}
\section{Discussion}
\label{sec:discussion}

Our results have a variety of additional implications for DM annihilation searches. \autoref{sec:future} discusses the prospects for improving the dwarf constraints with future UFD discoveries, while \autoref{sec:gce} considers implications of our results for the GCE.

\subsection{Prospects for Improving Dwarf Constraints}
\label{sec:future}

The known population of MW satellites has grown steadily with successive wide-field optical surveys---from the Sloan Digital Sky Survey through Pan-STARRS, the Dark Energy Survey, and ongoing programs such as DELVE and
UNIONS---and now numbers several tens of ultra-faint systems~\citep{Simon:2019nxf,2015ApJ...807...50B,DELVE:2024eud,2024ApJ...961...92S}. This census remains demonstrably incomplete, and the Vera C. Rubin Observatory's Legacy Survey of Space and Time~(LSST) is expected to discover many more UFDs over its survey lifetime~\citep{LSSTDarkMatterGroup:2019mwo}. Because UFDs are among the most DM-dominated systems known, and hence among the most promising targets for annihilation searches~\citep{Strigari:2007at}, the implications of this growing sample for indirect detection have been discussed in the literature: newly discovered satellites are routinely adopted as targets in gamma-ray analyses~\citep{Fermi-LAT:2015ycq,Rinchiuso:2019etv,Crnogorcevic:2023ijs}, and the constraining power anticipated from future discoveries has been forecast directly~\citep{Circiello:2024gpq}. How much these discoveries will tighten the annihilation limits, however, depends on whether they add nearby, high $J$-factor systems or only progressively fainter and more distant ones---equivalently, on whether the highest $J$-factor dwarfs have already been found.

\begin{figure}
    \includegraphics[width=0.98\columnwidth]{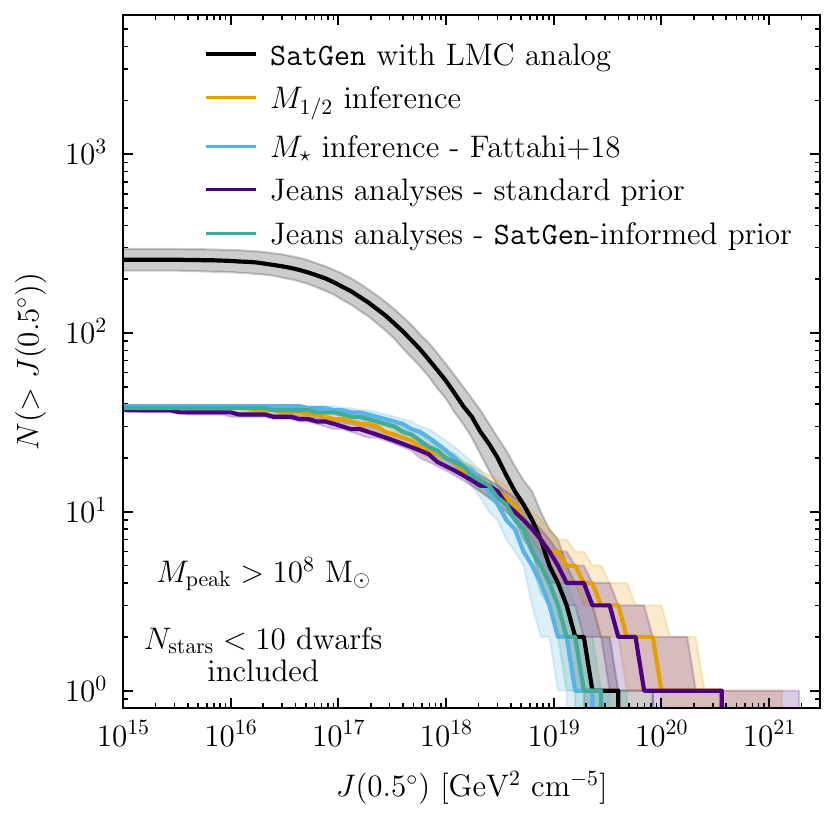}
    \caption{
    The median number of satellites with $J(0.5^{\circ})$ above a given value, along with 68\% containment bands. The results for the underlying Fiducial \SatGen{} model are shown in gray, selecting for LMC-associated hosts.  The results for the MW dwarf galaxies conditioned on \Mhalf{} weights are shown in light orange; those conditioned on \Mstar{} weights from the \Fattahi{} SHMR are shown in light blue.  Additionally, the distribution obtained using the standard-prior Jeans analyses is shown in indigo, while the distribution from Jeans analyses with the \SatGen{}-informed prior is shown in teal. We include the dwarf galaxies with $N_{\rm stars}<10$ (except for Tucana~V, see text) to compare to the indirect detection literature. The \SatGen{} results are not corrected for completeness and are thus an over-estimate of the expected distribution.
    }
    \label{fig:population_jfactors}
\end{figure}

\autoref{fig:population_jfactors} examines the dwarf galaxy $J$-factors at a population level. We compute the distribution of $N(>J(0.5^{\circ}))$ using the inferred $J$-factor PDFs.\footnote{This uses a similar procedure as described in \paperone{} for making the $\rho_{150}$ cumulative distributions.} Results are shown for the \Mhalf{}-inference method~(light orange) and \Mstar{} inference using the \Fattahi{} relation~(light blue).  For comparison, the results using the standard-prior and \SatGen{}-informed prior Jeans analyses are shown in indigo and teal, respectively. We include dwarf galaxies with $N_{\rm stars}<10$---with the exception of Tucana V, for which the Jeans analysis result is just the nonphysical prior---to compare to the literature on indirect detection. 

While the colored bands in \autoref{fig:population_jfactors} use a $J$-factor for each observed MW dwarf, the \SatGen{} expectation for the $J$-factor distribution is set through a different approach: we compute the $J$-factors of \SatGen{} halos at their cosmologically orbit-evolved locations, with the mock Earth-based observer placed in the Galactic plane 8~kpc from the Galactic center, sampling an azimuthal location within this circle randomly. The black line shows the median number of satellite halos (for hosts with an LMC analog,\footnote{As detailed in \paperone{}, we define an LMC analog as a satellite with $\Mpeak{}>10^{11}\;\Msun{}$ within 100 kpc of the \SatGen{} host center.} with that analog removed from the satellite sample) in \SatGen{} with a $J$-factor above a given value.  The corresponding 68\% containment band is shown in gray.   

For the median \SatGen{} host, the largest non-LMC $J$-factor is less than $\osim 3.5\times 10^{19}$~\Junits{}. Taking the \Mhalf{} weights (standard-prior Jeans analyses) at face value, the MW has $\osim3$~($2$) dwarf galaxies above this value. The \Mstar{}-inference results, on the other hand, are below this \SatGen{} expectation. We remark that these expectations are set without considering observational incompleteness effects and are therefore underestimates of the true $J$-factor functions that would be inferred from the full set of MW dwarf galaxies. Thus, if the \Mhalf{} or standard prior Jeans analyses represent the true $J$-factors, we expect to have already seen the highest $J$-factor dwarf galaxies. With the \SatGen{}-informed prior, we expect to see only one additional dwarf galaxy with a $J$-factor above $10^{19}\;\Junits{}$.

\subsection{Implications for Galactic Center Excess}
\label{sec:gce}

This subsection discusses the implications of the main results of this paper for the Fermi GCE~\citep{Goodenough:2009gk,Hooper:2010mq,Abazajian:2012pn,Fermi-LAT:2017opo}.  The box in the right panel of \autoref{fig:fermi_reweighting} 
encapsulates the best-fit cross sections from studies  attributing the GCE to gamma-ray flux from DM particles annihilating to $b\bar{b}$~\citep{Abazajian:2014fta, Calore:2014xka, Daylan:2014rsa,DiMauro:2021qcf}. 
The excess could also be an artifact of mismodeling or there could be an astrophysical contribution~\citep{Murgia:2020dzu}, which we highlight by extending the gray band in \autoref{fig:fermi_reweighting} (right panel) to zero.
Indeed, a variety of statistical and astrophysical modeling efforts find support for the hypothesis that the GCE is explained by an unresolved point-source population, which could be,  e.g., a population of dim millisecond pulsars~\citep{Abazajian:2010zy, Lee:2015fea, Linden:2016rcf, Chang:2019ars, Buschmann:2020adf, List:2020mzd, List:2021aer}. However, other studies find the point-source evidence to be sensitive to foreground modeling~\citep{Leane:2019xiy, List:2025qbx}. Multiple analyses have inferred a dimmer excess than the original papers with a morphology that traces the components of the stellar bulge~\citep{Macias:2016nev, Bartels:2017vsx, Macias:2019omb, Song:2024iup, Manconi:2025ogr}, which would disfavor the DM explanation. However, there is no consensus on the morphology issue with recent analyses using a variety of background models inferring both a GCE consistent with DM annihilation from a NFW-like density profile~\citep{Cholis:2021rpp, DiMauro:2026fnp} and a dimmer bulge-correlated GCE~\citep{Song:2024iup, Manconi:2025ogr}.

The \Mhalf{}-inferred limits exclude the thermal relic cross section for DM masses up to 146~\GeV{}, and at 40~\GeV{} (the mass relevant to a DM interpretation of the GCE) our \Mhalf{}-inferred limit reaches a factor of 4.7 below the thermal relic value. 
The standard-prior Jeans analysis limits exclude thermal relic cross section up to $M_{\chi}=127\;{\rm GeV}$ and by a factor of 4.2 at $M_{\chi}=40\;{\rm GeV}$
At the other end of the upper-limit envelope, the limits from the \Fattahi{}-weighted Jeans analysis exclude the thermal relic cross section for DM masses up to 67~\GeV{} and by a factor of 2.1 at 40~\GeV{}. 

Within the systematic uncertainties explored in this work, we conclude that the dwarf limits may not rule out the DM interpretation of the Fermi GCE. (Note that the best-fit cross sections for the GCE are consistent with cross sections smaller than the thermal relic cross section, see \autoref{fig:fermi_reweighting}.) This is especially true when considering that, as shown in \autoref{fig:Jfactors}, the most important halos are extended on the sky, which is an effect that was not accounted for in the~\citet{Circiello:2026inp} likelihoods that we use here; accounting for halo extension could even further weaken the dwarf constraints.  Moreover, as discussed in the previous subsection, future surveys are unlikely to find new UFDs that substantially increase the constraining power of the Fermi dwarf search. Given that Fermi is already well past its intended lifetime, we thus find it unlikely that Fermi analyses of the MW dwarf galaxies will be able to definitively rule out the DM interpretation of the GCE in the future. We note, however, that Fermi-LAT observations of the isotropic background and galaxy clusters have been used to set stronger constraints than the limits from the dwarfs~\citep{Delos:2023ipo,Crnogorcevic:2025nwp}, based on the finding that DM halos should contain highly concentrated low-mass cusps~\citep{Delos:2022bhp}.

\Needspace{4\baselineskip}
\section{Conclusions}
\label{sec:conclusions}

This work estimates the $J$-factors of the MW's dwarf galaxies using two complementary inference methods. First, we used \SatGen{} to generate large satellite populations and extracted a subsample that closely matched the properties of a given dwarf. We selected this subsample by matching either the observed galaxy's dynamical mass, \Mhalf{} (derived from its LOS velocity dispersion), or its stellar mass, \Mstar{} (assuming a SHMR relation). Second, we performed a Bayesian Jeans analysis using updated stellar kinematics from \citet{geha2026keckdeimosstellararchivei,geha2026keckdeimosstellararchiveii}. We evaluated this analysis across a sequence of priors that incorporated varying degrees of information, anchored by two limiting cases: a standard, wide log-uniform prior~\citep{Pace:2018tin} and a \SatGen{}-informed prior based on the dwarf's stellar mass and orbital location.  

As noted in \paperone{}, the most compact, lowest-luminosity UFDs have elevated velocity dispersions compared to CDM predictions, whereas the most diffuse UFDs have lower dispersions. While these deviations alter the inferred $J$-factors across the board, the high-$J$-factor dwarfs predominantly drive the final cross-section limits. As a result, DM annihilation limits derived from dwarf targets using the \Mhalf{}-inferred $J$-factors are $\osim 2\sigma$ stronger than expected by sampling from the \Mstar{}-inferred $J$-factor PDFs.
We have verified that these conclusions are robust to \SatGen{} modeling systematics such as different concentration-mass relations, SHMRs, \Mhalf{} estimators, host masses, LMC selection, and the \SatGen{} infall mass threshold (see also \paperone).  However, unaccounted observational systematics might still explain these results if they preferentially bias kinematic $J$-factor inferences upward. Unresolved binary star motion in UFDs without multiple epoch data is one such example~\citep{arroyopolonio2026estimatingdynamicalmassesdwarf}. Alternatively, the MW may simply be an atypical host.

Based on the comparison of $J$-factors derived from the distinct approaches used in this work, we arrive at three key conclusions. First, Jeans $J$-factors are highly sensitive to Bayesian prior assumptions. While standard procedures use wide, uninformative priors, this approach can drive inferred halo parameters toward unphysical values, particularly for dwarf galaxies with few measured stellar velocities. Adopting a \SatGen{}-informed prior resolves this by allowing the data to drive the results where kinematic constraints are strong, while defaulting to cosmological expectations where they are weak.

In detail, assuming the standard Jeans prior (which closely tracks the \Mhalf{}-inferred result), the 95\% confidence upper limit on the DM annihilation cross section to $b\bar{b}$ excludes the thermal relic cross section for DM masses, $M_\chi$, below 146~\GeV{}. At $M_\chi \sim 40~\GeV{}$---the mass most relevant for a $b\bar{b}$ interpretation of the Fermi GCE---this limit reaches a factor of 4.7 below the thermal relic value. The \SatGen{}-informed prior, however, only excludes the thermal relic cross section for masses below $M_\chi \sim 67$~\GeV{}, weakening the annihilation limit by a factor of 2--3 depending on DM mass. Given this impact, we advocate using an SHMR-weighted prior when computing upper bounds from Jeans-inferred $J$-factors. This approach prevents unphysical parameter regions from biasing the $J$-factors when kinematic data are sparse. Specifically, we recommend a prior based on the \Fattahi{} SHMR, which \paperone{} demonstrated provides the best fit to the dwarf galaxy kinematic data.

Second, the highest $J$-factor dwarf galaxies are extended compared to the Fermi PSF and should not be treated as point sources in analyses, as is typically done. For example, we find that Segue~1 and Draco have median 95\% containments of their $J$-factors of $\osim 1^{\circ}$, while the Fermi PSF at a photon energy of 10 GeV is $\osim 0.2^{\circ}$  These results motivate a new strategy where the joint posterior of the $J$-factor and the signal containment region is taken into account. It is possible that the constraints on the DM annihilation cross section may weaken significantly (by up to a factor of 2, see \citet{DiMauro:2022hue}) for some masses.

Finally, our analysis
of the $J$-factor population indicates that, given the current landscape of dwarf kinematics, the highest $J$-factor
MW satellites have likely already been discovered:
while next-generation surveys will continue to expand the UFD census, they
are unlikely to discover UFDs that will significantly strengthen these annihilation limits.

If, as our results suggest, the discovery of new UFDs is unlikely to significantly alter current annihilation limits, future progress will depend heavily on refining the density inferences of known targets. Addressing unresolved stellar systematics, adopting physically informed priors, and fully accounting for source extension will be essential to robustly probe the thermal relic cross section and continue pushing sensitivities even further.

\Needspace{4\baselineskip}
\section*{\textbf{Acknowledgments}}
We thank Yujin Park for collaboration during the early stages of this project, as well as Marla Geha, Andrew Pace, Nicholas Rodd, Daniel Hooper, and Alex Drlica-Wagner for useful discussions and comments. K.R. and B.R.S. are supported in part by the DOE award
DESC0025293. B.R.S. acknowledges support from the
Alfred P. Sloan Foundation and hospitality from the CERN theory group, where much of this work was completed.  M.L. and D.F. are supported by the Department of Energy~(DOE) under Award No. DE-SC0007968. M.L. is also supported by the Simons Investigator in Physics Award. M.K is supported by the National Science Foundation award PHY-2514888. This research used resources of the National Energy Research
Scientific Computing Center (NERSC), a U.S. Department of Energy Office of Science User Facility located
at Lawrence Berkeley National Laboratory, operated under Contract No. DE-AC02-05CH11231 using NERSC
award HEP-ERCAP0023978. This research has made use of the Keck Observatory Archive (KOA), which is operated by the W. M. Keck Observatory and the NASA Exoplanet Science Institute (NExScI), under contract with the National Aeronautics and Space Administration.

This report was prepared as an account of work sponsored by an agency of the United States Government. Neither the United States Government nor any agency thereof, nor any of their employees, makes any warranty, express or implied, or assumes any legal liability or responsibility for the accuracy, completeness, or usefulness of any information, apparatus, product, or process disclosed, or represents that its use would not infringe privately owned rights. Reference herein to any specific commercial product, process, or service by trade name, trademark, manufacturer, or otherwise does not necessarily constitute or imply its endorsement, recommendation, or favoring by the United States Government or any agency thereof. The views and opinions of authors expressed herein do not necessarily state or reflect those of the United States Government or any agency thereof.

\Needspace{4\baselineskip}
\bibliographystyle{aasjournal}
\bibliography{refs.bib}

\clearpage
\appendix
\renewcommand\thefigure{\thesection\arabic{figure}} 

\Needspace{4\baselineskip}
\section{Overview of Jeans Analyses}
\label{app:jeans}
\setcounter{figure}{0}
\setcounter{equation}{0}

To make contact between our semi-analytic inference procedure and standard literature methods, we implement a Jeans analysis~\citep{2008gady.book.....B}. For dispersion-supported systems, the spherical Jeans equation relates the density profile of the system to the velocity dispersion,
\begin{equation}
    \frac{d(\nu(r) \sigma_r^2(r))}{dr} + \frac{2\beta(r) \nu(r) \sigma_r^2(r)}{r} = -\frac{\nu(r)G M(r)}{r^2} \, ,
\end{equation}
where $\nu$ is the tracer density, $\sigma_{r,t}$ is the radial/tangential velocity dispersion, $\beta=1 - \sigma_t^2/\sigma_r^2$ is the velocity anisotropy, $G$ is the gravitational constant, and $M$ is the enclosed mass.  This equation has the integral solution
\begin{equation}
  \nu(r)\,\sigma_r^2(r)
  = \frac{1}{f(r)}
    \int_r^{\infty} f(s)\,\nu(s)\,\frac{G\,M(s)}{s^2}\,ds \, ,
  \label{eq:jeans-J1}
\end{equation}
where $f(r)=\exp{\left( 2 \int \frac{\beta(r')}{r'} dr' \right)}$. Projecting to the light of sight~(LOS),
\begin{equation}
  \Sigma(R)\,\sigmaLOS^2(R)
  = 2\int_R^{\infty}
      \left(1 - \beta\,\frac{R^2}{r^2}\right)
      \frac{\nu(r)\,\sigma_r^2(r)\,r}{\sqrt{r^2 - R^2}}\,dr \, ,
  \label{eq:projection-P1}
\end{equation}
where $\Sigma$ is the projected tracer density, $\sigmaLOS$ is the LOS velocity dispersion, and $R$ is the 2D distance from the center of the dwarf galaxy. As in~\citet{Pace:2018tin}, we assume a Plummer profile for the stellar density:
\begin{equation}
  \nu(r) = \frac{3}{4\pi r_p^3}\left(1 + \frac{r^2}{r_p^2}\right)^{-5/2}
  \qquad \text{and} \qquad 
  \Sigma(R) = \frac{1}{\pi r_p^2}\left(1 + \frac{R^2}{r_p^2}\right)^{-2} \, ,
  \label{eq:plummer}
\end{equation}
where the scale radius, $r_p$, is equivalent to the 2D projected half-light radius. For a real dwarf galaxy, the tracer density is elliptical, and we compute $r_p$ by azimuthally averaging the observed half-light radius, $r_p=a\sqrt{1-\epsilon}$, where $a$ is the semi-major axis of the half-light--containing ellipse and $\epsilon$ is the ellipticity. We additionally assume a constant velocity anisotropy, giving $f(r)=r^{2\beta}$, and parameterize the DM density by an NFW profile, yielding
\begin{equation}
  M(r) = 4\pi\,\rho_s\,r_s^3\;g(r/r_s),
  \qquad \text{where} \qquad
  g(x) = \ln(1+x) - \frac{x}{1+x} \, 
  \label{eq:nfw-mass}
\end{equation}
for the NFW scale density, $\rho_s$, and radius, $r_s$.  

We perform a Bayesian analysis to constrain the parameters $\left\{r_s, \rho_s, \bar V,  \beta\right\}$, where $\bar V$ is the systemic velocity, using the Gaussian unbinned log likelihood~\citep{Strigari:2007at}
\begin{equation}
\ln \mathcal{L} = \sum_i \ln \mathcal{L}_i, \qquad \text{where} \quad 
  \ln\mathcal{L}_i
  = -\tfrac{1}{2}\ln\!\bigl(2\pi s_i^2\bigr)
    - \frac{(V_i - \bar V)^2}{2\,s_i^2}
    \quad \text{and} \quad 
    s_i^2 \equiv \sigmaLOS^2(R_i;\,\rho_s, r_s, \beta, r_p)
              + \sigma_{\varepsilon,i}^2 \,.
  \label{eq:per-star-lnlike}
\end{equation}
Here, the sum is taken over a dwarf's member stars, $V_i$ is the stellar LOS velocity, $\sigma_{\varepsilon,i}^2$ is the corresponding measurement uncertainty, and $R_i$ is the distance of the star from the center of the dwarf galaxy. We use log-uniform priors on $r_s$ and $\rho_s$, $-2<\log_{10} {r_s}/{\rm kpc}<1$ and $4<\log_{10} {\rho_s}/{{\rm \Msun{}~kpc}^{-3}}<14$, a uniform prior on a symmetrized anisotropy parameter $\Tilde{\beta}={\beta}/{(2-\beta)}$, $-0.95<\Tilde{\beta}<1$, and a uniform prior on $\bar V$, $\bar V_0 - 10~\kms{}<\bar V<\bar V_0 + 10~\kms{}$, where $\bar V_0$ is the inverse-variance weighted mean of the stellar LOS velocity measurements. We additionally impose a prior of $r_s>r_{1/2}$ as in~\citet{Pace:2018tin}. To profile over the uncertainty in $r_p$, we sample from a Gaussian prior determined by the observational uncertainties for $a$ and $\epsilon$. 

Finally, we correct for perspective motion effects as outlined in \citet{Kaplinghat:2008sm,Walker:2008ji}. Away from the center of an extended dwarf galaxy, the local LOS direction is rotated relative to the LOS through the galaxy center. Thus, the perceived stellar LOS velocity and the velocity component parallel to the systemic LOS velocity differ; in particular, this shift scales with the bulk motion of the dwarf galaxy and the angle from the dwarf galaxy center. We correct for this shift by adjusting the measured LOS velocity for each star in each dwarf galaxy by the projection of the galaxy's proper motion onto the stellar LOS---the small-angle approximation allows us to drop an additional contribution from the systemic LOS velocity. The velocity shift in \kms{} is
\begin{equation}
\label{eq:persp}
\Delta V_\mathrm{persp}(\alpha,\delta) = A\,d\,\bigl(\mu_{\alpha}\cos{\delta}\times\Delta\alpha\cos\delta + \mu_\delta\times\Delta\delta\bigr) \, ,
\end{equation}
where $A = 4.74047~\mathrm{km\,s^{-1}}\,(\mathrm{mas\,yr^{-1}\,kpc})^{-1}$ is a unit conversion factor,
$d$ is the observed dwarf galaxy distance from Earth in kpc, $\mu_\alpha\cos\delta$ and $\mu_\delta$ are
the proper motion components in $\mathrm{mas\,yr^{-1}}$ ($\delta$ is the dwarf galaxy declination),
and $\Delta\alpha$ and $\Delta\delta$ are right ascension and declination angular offsets between the star and dwarf galaxy center in radians. The perspective-corrected LOS velocities that enter the likelihood in \autoref{eq:per-star-lnlike} are $V_i=V_{i,0} -\Delta V_{\rm persp}$, where $V_{i,0}$ is the original measured stellar LOS velocity. We apply this correction for every dwarf galaxy in the sample. As with the other dwarf galaxy observables, we profile over uncertainties in $\mu_{\alpha}$ and $\mu_{\delta}$ by sampling from a Gaussian prior determined by the observational uncertainties.

All dwarf galaxy structural and systemic parameters ($a$, $\epsilon$, $\mu_{\alpha}$, $\mu_{\delta}$, $\alpha$, $\delta$, $d$) are taken from the LVDB as described in \paperone. Stellar velocity data are taken from the KDSA~\citep{geha2026keckdeimosstellararchivei,geha2026keckdeimosstellararchiveii}, where available, and from the works corresponding to the listed LVDB velocity dispersions otherwise~\citep{Walker:2008ax,Kirby:2015ija,Koposov:2015jla,Walker:2015mla,DES:2016vji,Koposov_2018,MagLiteS:2018ylp,DES:2019ncb,Ji_2021,Chiti:2022pco,Bruce:2023qwg,Chiti_2023,hansen2024chemicaldiversitysmallscales,DELVE:2023xch,DELVE:2024yct}. Where provided by the authors, we use precompiled tables of per-star velocities and uncertainties. Otherwise, when per-star precompiled results are unavailable, we compile measurements for a given star from $N$ measurement epochs into a single LOS velocity and measurement uncertainty as follows:
\begin{equation}
  V_{i,0}
  = \frac{\sum_{j=1}^{N} v_j / \sigma_{{\rm stat},j}^2}
         {\sum_{j=1}^{N} 1/\sigma_{{\rm stat},j}^2} \,,
  \qquad
  \sigma_{\epsilon, i}
  = \sqrt{\frac{1}{\sum_{j=1}^{N} 1/\sigma_{{\rm stat},j}^2}
          + \sigma_\mathrm{sys}^2} \, ,
  \label{eq:ivw}
\end{equation}
where the $v_j$ are individual measurements, the $\sigma_{{\rm stat},j}$ are statistical uncertainties, and $\sigma_{\rm sys}$ is a systematic uncertainty. Since each dataset reports stellar measurement velocities with systematic uncertainties already added in quadrature, we treat all quoted uncertainties as statistical and set $\sigma_{\rm sys}=0$. This multi-epoch measurement combination procedure is required for seven dwarf galaxies: Carina~II, Carina~III, Eridanus~II, Grus~I, Tucana~II, Tucana~IV, and Tucana~V. We note that for Tucana II, the dataset is a compilation of measurements from multiple instruments---MIKE, IMACS, M2FS, and MagE~\citep{2003SPIE.4841.1694B,Marshall:2008de,Dressler_2011,2012SPIE.8446E..4YM}---which have systematic velocity measurement offsets relative to each other. We account for these offsets relative to MIKE (0.0~\kms{}, $-2.2$~\kms{}, $-2.5$~\kms{}, and $-1.0$~\kms{}, respectively, as documented in \citet{Chiti_2023}) in combining multi-epoch measurements. Additionally, to flag possible binaries, we compute a $\chi^2$ statistic,
\begin{equation}
    \chi^2_i = \sum_{j=1}^N \frac{(v_j - V_i)^2}{\sigma_{{\rm stat}, j}^2}
\end{equation}
and flag a star as ``variable'' if $p(\chi^2_i)<0.01$ under a $\chi^2_{N-1}$ distribution. We additionally retain ``variability'' or ``binary'' flags preexisting within the stellar datasets.

After initial data processing, we follow the selection procedure taken by \citet{geha2026keckdeimosstellararchiveii} in their velocity dispersion computations and make the following selection for stars that enter the Jeans analyses: we discard flagged variable or binary stars, stars with a probability of membership below 0.5, and stars outside twice the azimuthally-averaged 3D half-light radius (i.e., twice the $r_{1/2}$ listed in \autoref{tab:J_meta}). For the dwarf galaxies in the dataset of \citet{geha2026keckdeimosstellararchivei}, the variability and membership selections are precomputed, and we simply select on the provided $P_{\rm mem\_novar}$ flag. We apply this procedure uniformly across the dwarf galaxy samples without making further selections.

\begin{figure*}[p]
    \includegraphics[width=0.98\textwidth]{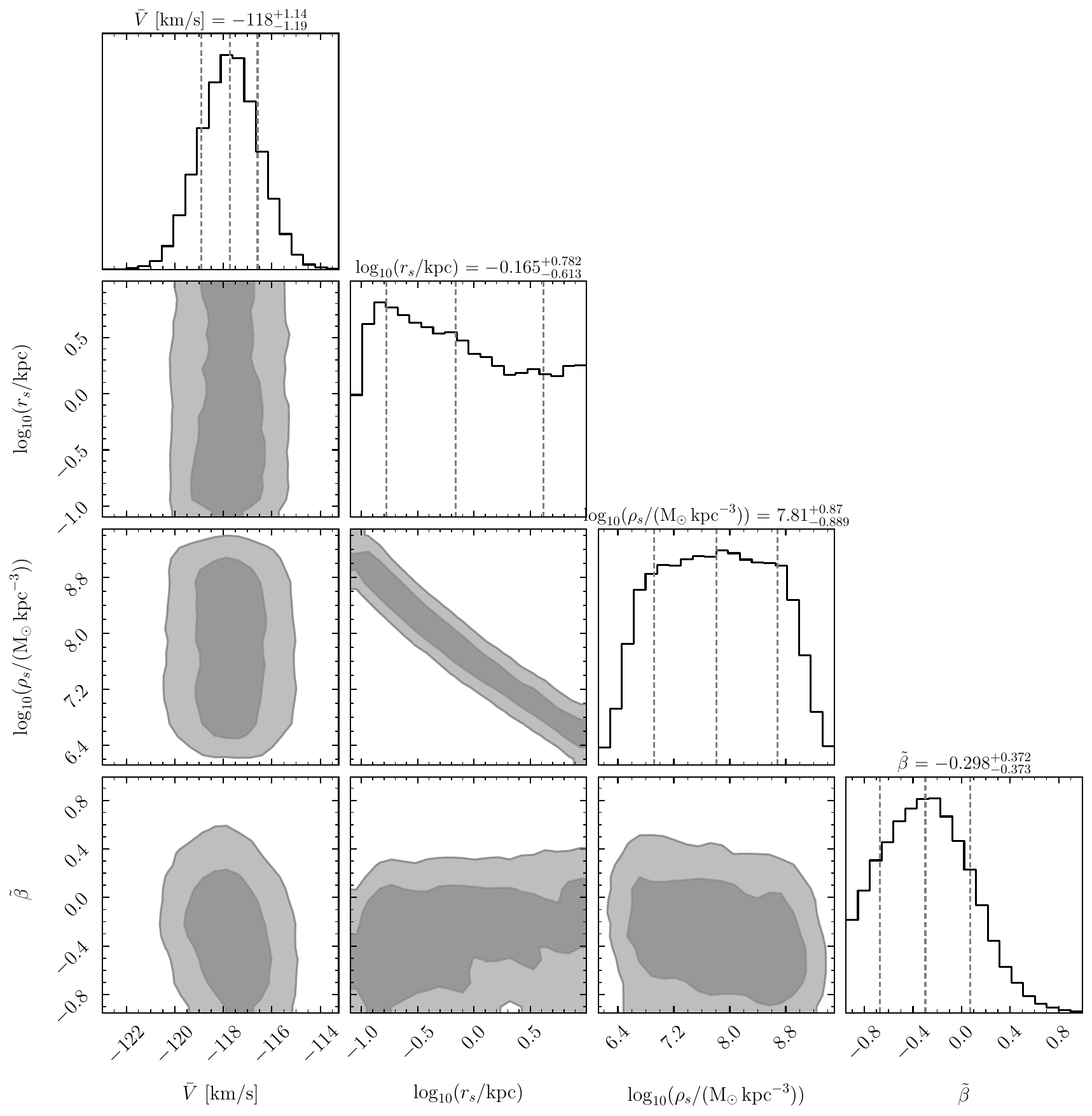}
    \caption{Jeans analysis posteriors on $\bar V$, $r_s$, $\rho_s$, and $\Tilde{\beta}$ for Ursa~Major~II.}
    \label{fig:jeans_posterior}
\end{figure*}

As an example, \autoref{fig:jeans_posterior} provides the Jeans analysis posteriors for $r_s$, $\rho_s$, $\bar V$, and $\Tilde{\beta}$ for Ursa~Major~II. As expected from the degeneracy between $\rho_s$ and $r_s$ in the NFW expression for enclosed mass, neither $r_s$ nor $\rho_s$ are independently constrained. From these posteriors, one can calculate the astrophysical $J$-factor, as described in \autoref{sec:jfactor}. To profile over distance uncertainties, we sample from a Gaussian prior determined by the observational uncertainties in the dwarf galaxy's distance. 

\Needspace{4\baselineskip}
\section{Full Set of $J$-factor Priors}
\label{app:all_priors}
\setcounter{figure}{0}

\begin{figure*}[p]
    \includegraphics[width=0.98\textwidth]{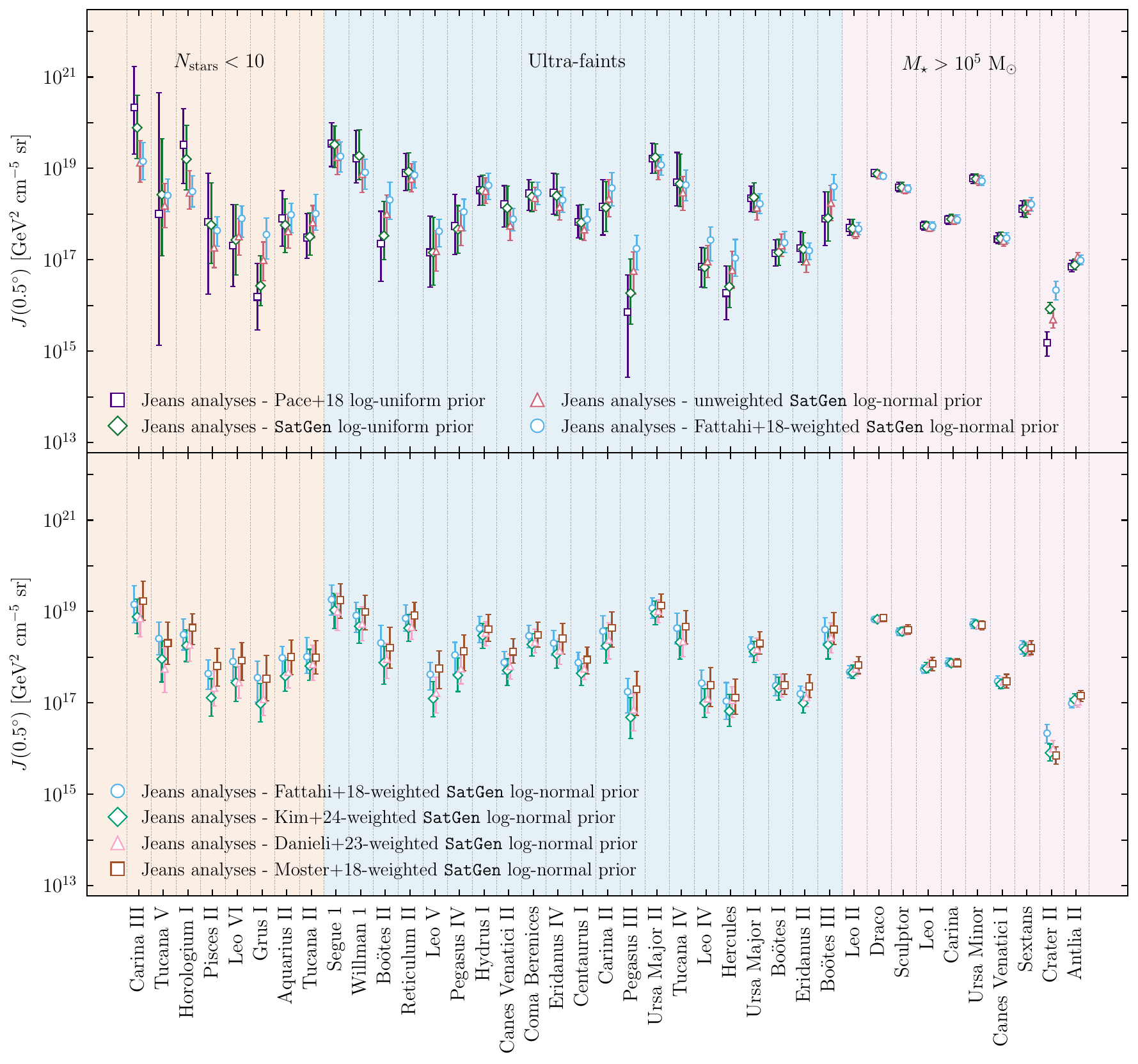}
    \caption{A comparison of inferred astrophysical $J$-factors for dwarf galaxies under various Jeans analysis priors, separated into the three nominal categories and ordered by increasing $r_{1/2}$ within each category.  
    Top panel: The indigo squares show the Jeans results using the wide log-uniform priors from~\citet{Pace:2018tin}. The dark green diamonds and rose triangles show the Jeans results for a \SatGen{}-motivated log-uniform and log-normal prior, respectively (see text for details). Finally, for the light blue circles, we repeat the same process using the \Fattahi{}-weighted and distance-selected \SatGen{} distributions. For dwarf galaxies with larger kinematic uncertainties, there exist up to order of magnitude differences between Jeans analysis results using \SatGen{}-driven informative priors and results with log-uniform priors. By contrast, for the dwarf galaxies with $\Mstar>10^5~\Msun{}$, with the exception of Antlia~II and Crater~II, constraining data leads to limited prior dependence in the inference results.
    Bottom panel: Same as \autoref{fig:jeans_panel}, except for Jeans results with SHMR-weighted priors. The light blue circles, light green diamonds, pink triangles, and brown squares indicate results with \Fattahi{}, \Kim{}, \Danieli{}, and \Moster{}, respectively. The results with the \Kim{} and \Danieli{} priors are consistently smaller than the results with the \Fattahi{} and \Moster{} priors by up to 0.5~dex across the $N_{\rm stars}<10$ dwarfs and Ultra-faints. The results for the $\Mstar>10^5~\Msun{}$ dwarfs show little prior dependence, except for Antlia~II and Crater~II.
    }
    \label{fig:jeans_panel}
\end{figure*}

This appendix compares the astrophysical $J$-factors and 95\% upper limits on $\chi\chi \to b\bar{b}$ across every prior considered in this work. The top panel of \autoref{fig:jeans_panel} shows the inferred astrophysical $J$-factor across the four priors discussed in \autoref{sec:jeans}: the wide log-uniform priors from \citet{Pace:2018tin} (indigo squares, the standard prior from the main text), the \SatGen{}-motivated log-uniform prior (dark green diamonds), the \SatGen{}-motivated log-normal prior (rose triangles), and the \Fattahi{}-weighted and distance-selected \SatGen{} distributions (light blue circles, the \SatGen{}-informed prior from the main text). More informative priors yield more constrained $J$-factors for the ultra-faints, but have minimal effects on the $\Mstar{}>10^5\;\Msun{}$ $J$-factors, with the exception of Crater~II and Antlia~II. 

The bottom panel shows the inferred astrophysical $J$-factor across variations to the specific choice of SHMR used for the \SatGen{}-informed prior: \Fattahi{}, \Kim{}, \Danieli{}, and \Moster{} in light blue circles, light green diamonds, pink triangles, and brown squares, respectively. For the ultrafaints, results with the \Kim{} and \Danieli{} priors are broadly smaller than the \Fattahi{} and \Moster{} results by $0.2\text{--}0.5$~dex since the \Kim{} and \Danieli{} SHMRs pick out lower-mass halos at the same \Mstar{}.

\begin{figure*}
    \centering
    \includegraphics[width=0.49\textwidth]{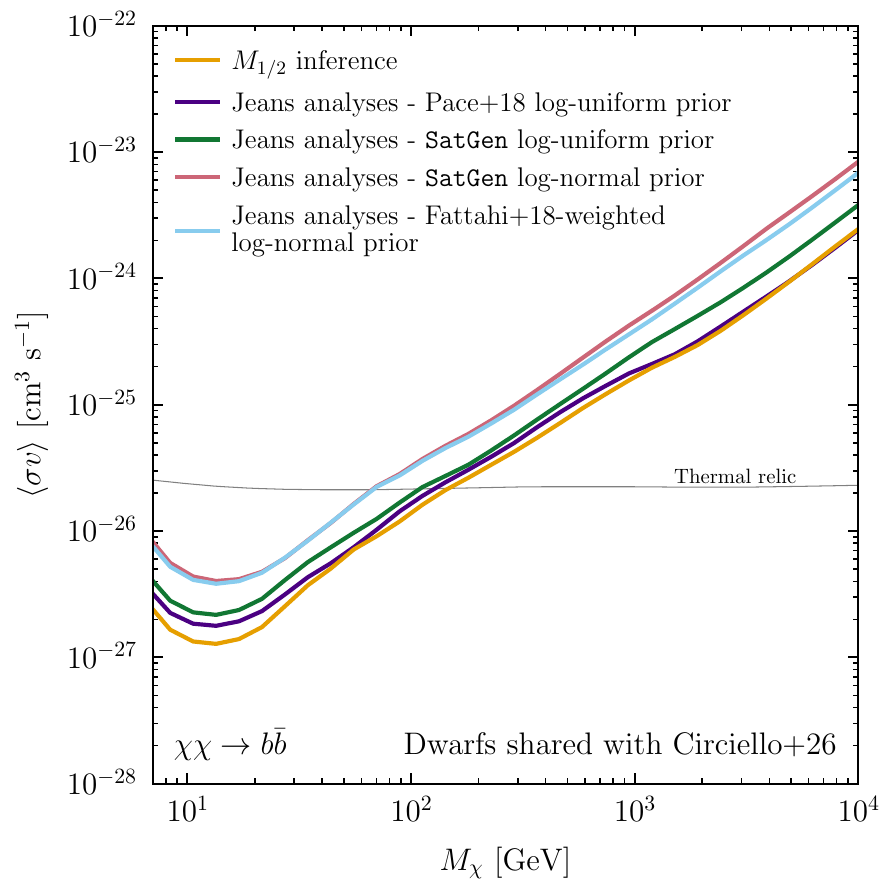}
    \includegraphics[width=0.49\textwidth]{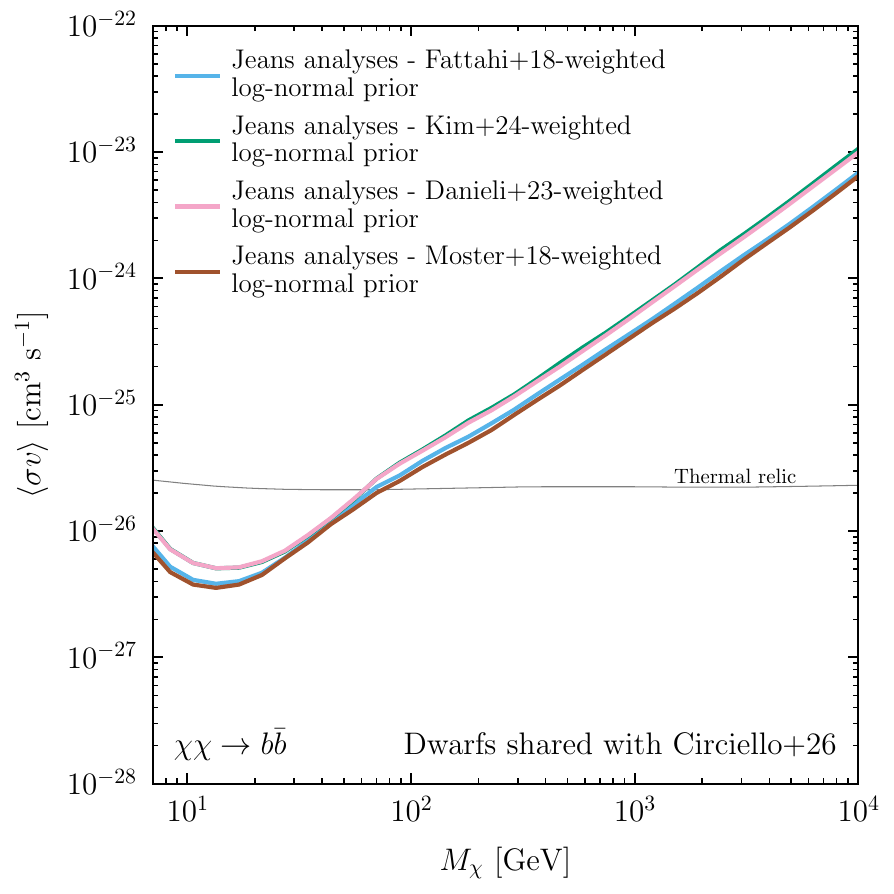}
    \caption{Left: The 95\% upper limits on the $\chi\chi\to b \bar{b}$ velocity-averaged DM annihilation cross section with five different sets of $J$-factors, using the sample of 30 dwarf galaxies in common between this work and the benchmark sample from \citet{Circiello:2026inp}. The light orange line shows the \Mhalf{} inference result and the indigo line shows the Jeans analysis result using the log-uniform prior on $r_s$ and $\rho_s$ from \citet{Pace:2018tin}. The dark green line shows the Jeans analysis result with a log-uniform prior restricted to the region of \SatGen{} support in $(\rho_s, r_s)$ space. The rose and light blue lines show the results using two ``physics-informed" priors that are fit to distributions of \SatGen{} halos---one with log-normal distributions fit to unweighted \SatGen{} halos from the Fiducial run (rose), and one following the same process with \Fattahi{}-weighted \SatGen{} halos and including the dwarf-galaxy specific galactocentric distance selections. The limits with these two physics-informed priors are weaker than the \Mhalf{} and Jeans results by a factor of $\osim 2\text{--}3$ across the DM mass range. See \autoref{app:limit_comparison} for a comparison to other published limits in the literature. Right: Same as the right, but comparing Jeans results with four different SHMR-weighted priors: \Fattahi{} (light blue), \Kim{} (light green), \Danieli{} (pink), and \Moster{} (brown). All results include the galactocentric distance selection for each dwarf galaxy. The limits are within a factor of 1.4 of each other at all masses.}
    \label{fig:limit_all_priors}
\end{figure*}

The left and right panels of \autoref{fig:limit_all_priors} show the 95\% upper limits on $\chi\chi \to b \bar{b}$ with the same priors and color scheme as the top and bottom panels of \autoref{fig:jeans_panel}, respectively. We additionally show the \Mhalf{} inference results in the left panel in light orange. Using $J$-factors from a Jeans analysis with a \SatGen{}-informed prior weakens the limit by up to a factor of 3, while the differences in the result between different SHMRs are within a factor of 1.4 at all DM masses.

\Needspace{4\baselineskip}
\section{Crater~II and Antlia~II structural parameters}
\label{app:craterantlia}
\setcounter{figure}{0}

\begin{figure*}
    \includegraphics[width=0.49\textwidth]{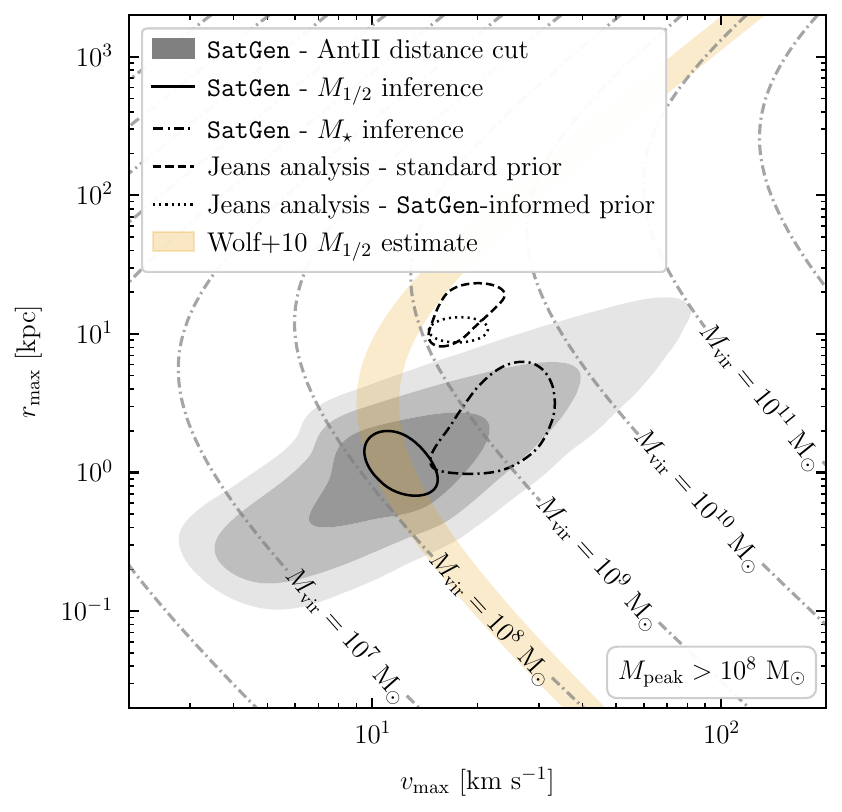}
    \includegraphics[width=0.49\textwidth]{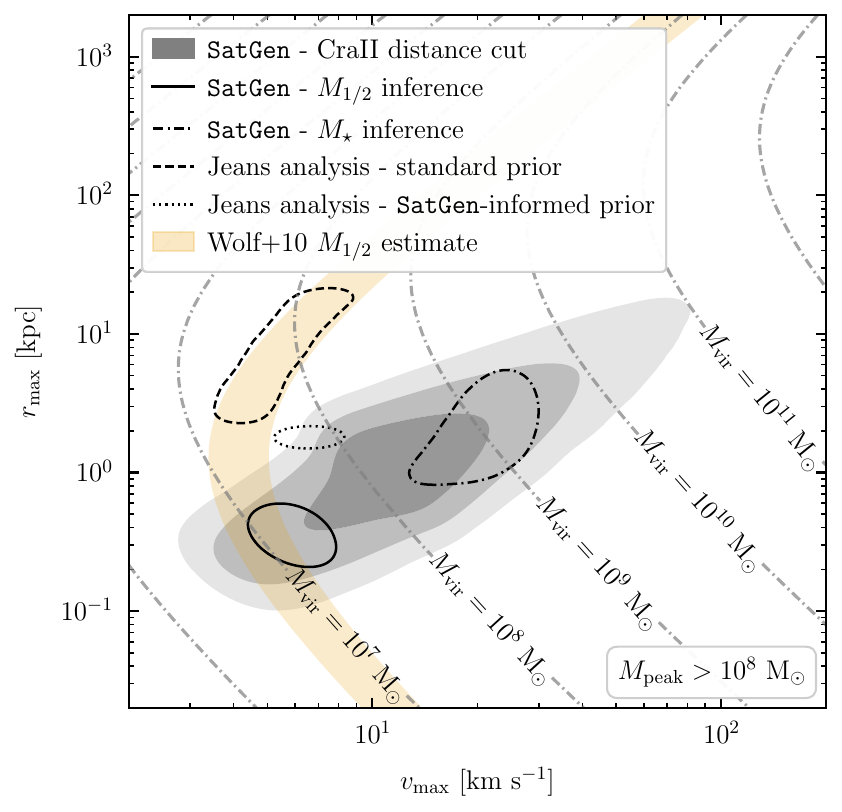}
    \caption{
    Inferred values for the $z=0$ maximum circular velocity \vmax{} and the radius \rmax{} at which it occurs for the MW dwarfs Antlia~II~(AntII, left) and Crater~II~(CraII, right). In gray are the 68, 95, and 99.5\% containment regions for the \SatGen{} satellite halos satisfying the distance selections for the respective dwarf galaxies.
    The solid and dash-dotted black contours show the 68\% containment for the profile parameters inferred using \Mhalf{} and \Mstar{}, respectively. The \Mstar{} inference assumes the SHMR from \Fattahi{}. Both the \Mhalf{}- and \Mstar{}-inference results are computed using the Fiducial \SatGen{} run and respective dwarf galaxy distance selections. The dashed- and dotted-black contours shows the corresponding region using the results from the Jeans analysis with the standard and \SatGen{}-informed prior, respectively.
    The light orange band shows the locus of NFW halo parameters whose enclosed mass at the half-light radius of the respective dwarf galaxies falls within the 68\% confidence interval of the observed \Mhalf{} (as computed using the \Wolf{} estimator).
    The gray dot-dashed contours denote lines of constant \Mvir{}---the virial mass of a NFW halo with the given parameters at $z=0$.
    For both dwarf galaxies, the standard prior Jeans analysis posteriors lie outside the 99.5\% containment of the \SatGen{} distribution.
    }
    \label{fig:rmaxvmax_diffuse}
\end{figure*}

In \autoref{sec:jfactor}, we infer dwarf galaxy $J$-factors using Jeans analyses and the \Mhalf{}- and \Mstar{}-inferences.
We find that the \Mhalf{}-inference and Jeans results are broadly consistent with each other, except for Crater~II and Antlia~II, where the latter method leads to $J$-factors that are 1--2 orders of magnitude discrepant. This subsection explores how the \Mhalf{}-inference and Jeans analyses differ at the level of the inferred structural parameters. We find that Jeans analyses with wide priors pick out halos outside the 99.5\% containment of the \SatGen{} distribution for these two dwarf galaxies.

\autoref{fig:rmaxvmax_diffuse} shows the 68\% containment contours in $(v_{\rm max}, r_{\rm max})$ of the \Mhalf{} inference (solid black), \Fattahi{} \Mstar{}-inference (dash-dotted black), and Jeans analyses with the standard prior (dashed black) and \SatGen{}-informed prior (dotted black). The \SatGen{} distribution with the dwarf galaxy specfic galactocentric distance selection is in shaded gray. We show the locus of NFW halo parameters whose $M_{1/2}$ falls within the 68\% confidence interval of the \Wolf{} estimator in light orange. Results for Antlia~II are shown in the left panel, while those for Crater~II are shown on the right. 

For both Crater~II and Antlia~II, the standard Jeans likelihood prefers a combination of halo parameters outside the 99.5\% containment of  the corresponding \SatGen{} distribution; the prior used in these analyses allows for these halo parameters. The resulting halo profiles are therefore inconsistent with the $\Mhalf{}$- and $\Mstar{}$-inference results. The \SatGen{}-informed Jeans analysis result for Antlia~II also lies outside of the 99.5\% containment of the \SatGen{} distribution. For Crater~II, this result lies within the 99.5\% containment, but is still disjoint with the \Mhalf{}- and \Mstar{}-inference results.

Additionally, while the Jeans result for Crater~II is consistent with the \Wolf{} \Mhalf{} band, that for Antlia~II is not. This is likely because Antlia~II has a velocity gradient that is modeled in the \sigmaLOS{} fit by \citet{Ji_2021}---the source of the LVDB value---but not in our Jeans analysis. \citet{Ji_2021} find that not including the velocity gradient inflates the best-fit velocity dispersion by $\osim1.5$~\kms{}. This would push the Jeans contour to higher virial masses, as observed. Since a consistent incorporation of a velocity gradient would require dropping the assumption of spherical symmetry in the Jeans analyses, as in e.g., axisymmetric Jeans modeling~\citep{Cappellari:2008kd}, we leave a consistent incorporation of velocity gradients to future work.

\Needspace{4\baselineskip}
\section{Comparison to Previous Literature Results}
\label{app:limit_comparison}
\setcounter{figure}{0}

\begin{figure*}
    \includegraphics[width=0.98\textwidth]{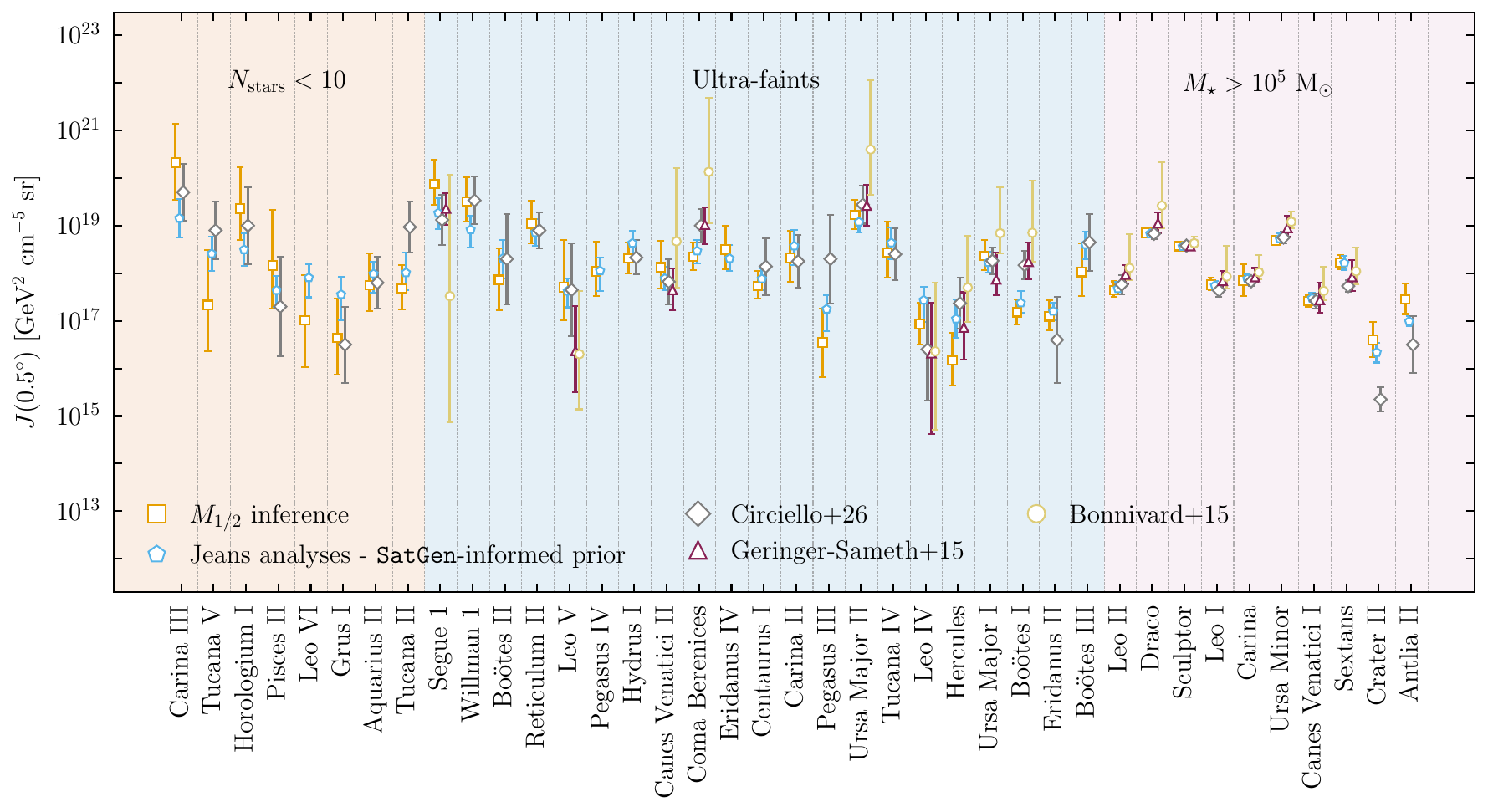}
    \caption{Comparison of $J$-factors from the \Mhalf{} inference  (light orange squares) and Jeans analysis with the \SatGen{}-informed prior (light blue pentagons) to previous literature $J$-factors, where available. The grey diamonds indicate $J$-factors from the compilation of \citet{Circiello:2026inp}. For Sculptor, which is left out of \citet{Circiello:2026inp} for data quality reasons, we show the result from \citet{Pace:2018tin}---the source from which the majority of the \citet{Circiello:2026inp} $J$-factors are taken. The dark magenta triangles indicate $J$-factors from \citet{Geringer-Sameth:2014yza}, and the tan circles indicate $J$-factors from \citet{Bonnivard:2015xpq}.}
    \label{fig:literature_comparison}
\end{figure*}

\begin{figure*}
    \centering
    \includegraphics[width=0.49\textwidth]{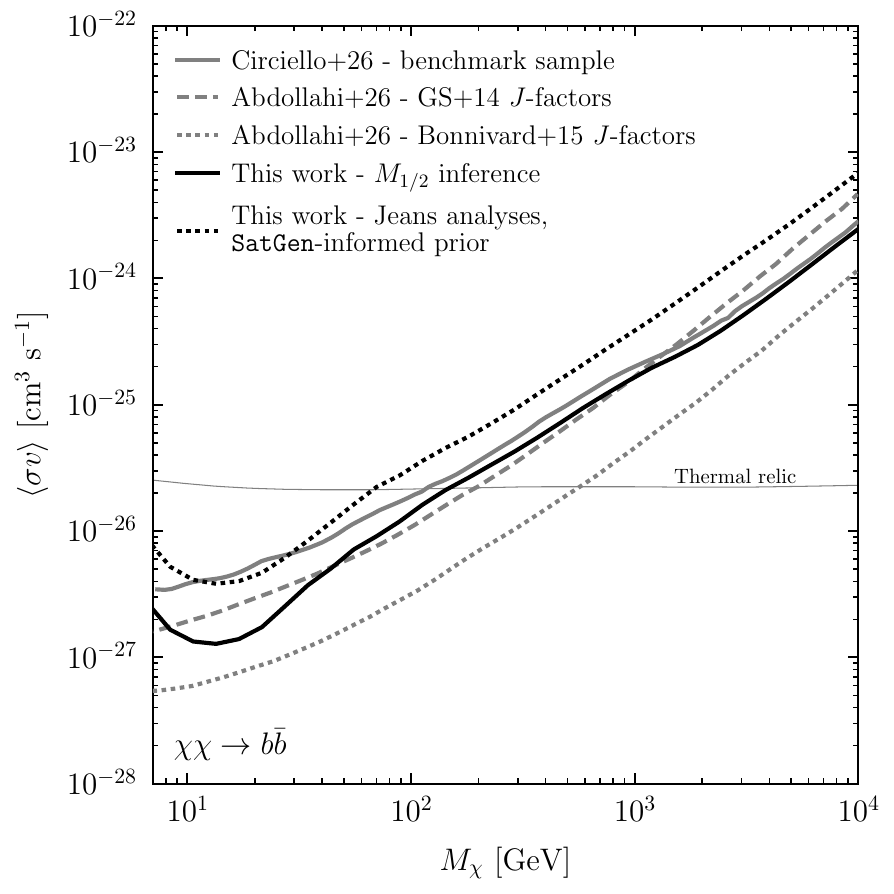}
    \caption{
    Comparison of the two upper limits on $\chi\chi\to b \bar{b}$ that bracket our results from \autoref{fig:fermi_reweighting} (\Mhalf{} inference, dashed black, and Jeans analysis with \SatGen{}-informed prior, dotted black) to recent literature results. These two limits recast the flux upper limits from \citet{Circiello:2026inp} for the same 30 dwarf galaxies as in \autoref{fig:fermi_reweighting}. The solid gray line shows the published limit from \citet{Circiello:2026inp}, who update the results from \citet{McDaniel:2023bju} using the same 42 dwarf galaxies and $J$-factors. We additionally show two limits from the Fermi-LAT Collaboration, presented by \citet{Fermi-LAT:2025gei}. These results use 18 dwarf galaxies and consider two sets of $J$-factors. Results with the $J$-factors of \citet{Geringer-Sameth:2014yza} are shown in dashed gray, while results with the $J$-factors of \citet{Bonnivard:2015xpq} are shown in dotted gray. 
    }
    \label{fig:limit_comparison}
\end{figure*}

This appendix compares our astrophysical $J$-factors and 95\% upper limits on $\chi\chi \to b\bar{b}$ to recent literature results. The $J$-factor comparison is shown in \autoref{fig:literature_comparison}, which compares the \Mhalf{}-inferred results (light orange squares) and Jeans results for the \SatGen{}-informed prior (light blue pentagons) to various literature results: \citet{Circiello:2026inp} in grey diamonds, \citet{Geringer-Sameth:2014yza} in dark magenta triangles, and \citet{Bonnivard:2015xpq} in tan circles, where available.

The astrophysical $J$-factors shown in \autoref{fig:literature_comparison} translate into the limits shown in \autoref{fig:limit_comparison}. For comparison, this figure reproduces two limits shown in \autoref{fig:fermi_reweighting}: one using the \Mhalf{}-inferred $J$-factors (solid black) and one using the Jeans-inferred $J$-factors with \SatGen{}-informed priors (dotted black).  The solid gray line shows the published limit from \citet{Circiello:2026inp}, who update the results of \citet{McDaniel:2023bju} using the same sample of 42 dwarf galaxies and the same $J$-factors. The differences between the black and gray results are driven by the use of different $J$-factors and dwarf galaxies.

\autoref{fig:limit_comparison} also shows results from recent work by the Fermi-LAT Collaboration~\citep{Fermi-LAT:2025gei}. That work used 18 dwarf galaxies and presented upper limits using two different sets of $J$-factors. The results using $J$-factors from \citet{Geringer-Sameth:2014yza} are shown in dashed gray. The analysis procedure used to obtain these $J$-factors contains three key differences with respect to our procedure: they allow for non-NFW profiles, truncate the $J$-factor computation at the radius of the outermost observed star in each dwarf galaxy, and use different stellar membership selections.  The results using the $J$-factors from \citet{Bonnivard:2015xpq} are shown in dotted gray. This work also allows for non-NFW profiles and uses different stellar membership selections. We also emphasize that \citet{Fermi-LAT:2025gei} use a different data analysis procedure than \citet{Circiello:2026inp}---and therefore our recast results.

\end{document}